\documentclass[aps,prb,twocolumn,groupedaddress]{revtex4-1}
\usepackage{graphicx}
\usepackage{float}
\usepackage{amsmath,amsfonts,amssymb,mathtools}
\usepackage{multirow}

\begin{document}



\title{Ephemeral data derived potentials for random structure search}

\author{Chris J.\ Pickard} \email[]{cjp20@cam.ac.uk}
\affiliation{Department of Materials Science \& Metallurgy, University
  of Cambridge, 27 Charles Babbage Road, Cambridge CB3~0FS, United
  Kingdom} \affiliation{Advanced Institute for Materials Research,
  Tohoku University 2-1-1 Katahira, Aoba, Sendai, 980-8577, Japan}


\date{\today}

\begin{abstract}
  Structure prediction has become a key task of the modern atomistic
  sciences, and depends on the rapid and reliable computation of
  energy landscapes. First principles density functional based
  calculations are highly reliable, faithfully describing entire
  energy landscapes.  They are, however, computationally intensive and
  slow compared to interatomic potentials. Great progress has been
  made in the development of machine learning, or data derived,
  potentials, which promise to describe entire energy landscapes at
  first principles quality. Compared to first principles approaches,
  their preparation can be time consuming and delay searching. Ab
  initio random structure searching (AIRSS) is a straightforward and
  powerful approach to structure prediction, based on the stochastic
  generation of sensible initial structures, and their repeated local
  optimisation. Here, a scheme, compatible with AIRSS, for the rapid
  construction of disposable, or ephemeral, data derived potentials
  (EDDPs) is described. These potentials are constructed using a
  homogeneous, separable manybody environment vector, and iterative
  neural network fits, sparsely combined through non-negative least
  squares. The approach is first tested on methane, boron nitride,
  elemental boron and urea. In the case of boron, an EDDP generated
  using data from small unit cells is used to rediscover the complex
  $\gamma$-boron structure without recourse to symmetry or
  fragments. Finally, an EDDP generated for silane (SiH$_4$) at 500
  GPa enables the discovery of an extremely complex, dense, structure
  which significantly modifies silane's high pressure phase
  diagram. This has implications for the theoretical exploration for
  high temperature superconductivity in the dense hydrides, which have
  so far largely depended on searches in smaller unit cells.
\end{abstract}


\maketitle

\section{Introduction}


The knowledge of the arrangement, and nature, of atoms in a system is
an essential starting point for its theoretical or computational
study. First-principles approaches to crystal structure prediction
have provided a route to this knowledge which is independent of
experiment or intuition.\cite{oganov2019structure} Early approaches
were based on evolutionary algorithms,\cite{oganov2006crystal} or
random search,\cite{pickard2006high,pickard2011ab} but many related
algorithms have been proposed
since.\cite{lonie2011xtalopt,wang2012calypso} Over the last decade and
a half, first-principles structure prediction has led to a number of
computational “discoveries”. These include dense transparent
sodium,\cite{ma2009transparent} the structure of phase III of hydrogen
and its mixed phase IV,\cite{pickard2007structure} and complex
host-guest structures in aluminium at terapascal
pressures.\cite{pickard2010aluminium} The first application of random
structure search\cite{pickard2006high} was to testing Ashcroft’s
prediction\cite{ashcroft2004hydrogen} that compressed hydrides might
offer a route to high temperature
superconductivity.\cite{pickard2020superconducting} This has been
dramatically confirmed with the experimental discovery of
superconductivity in hydrogen sulphide at
203K\cite{drozdov2015conventional} and 250K in
LaH$_{10}$.\cite{drozdov2019superconductivity} In both cases the
structures were predicted from first principles and the
superconductivity anticipated
computationally.\cite{duan2014pressure,peng2017hydrogen,liu2017potential}

\begin{figure}[]
\centering
\includegraphics[width=0.45\textwidth]{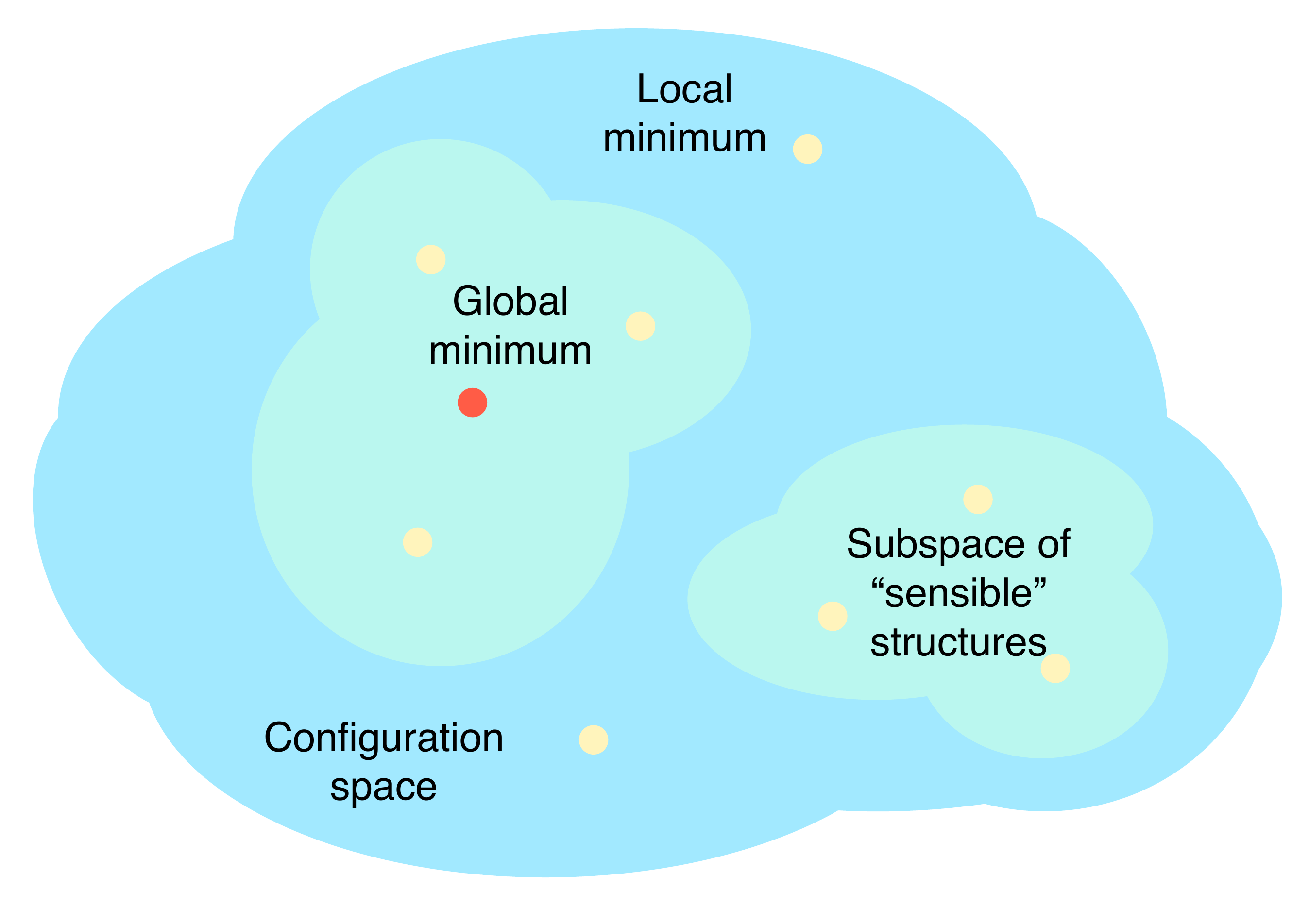}
\caption{A sketch of configuration space, highlighting regions which
  may be reached starting from structures assembled according to
  physically motivated biases and/or constraints. In general, the
  volume of configuration space accessible from these ``sensible''
  initial structures will be very small compared to the total volume
  of configuration space.}\label{figure_landscape}
\end{figure}

Ab initio random structure searching (AIRSS) is a particularly simple,
yet powerful, approach to structure prediction.\cite{pickard2011ab}
Random structures are generated and relaxed to nearby local minima of
the energy landscape, repeatedly and in parallel. With a focus on
exploration rather than exploitation, the initial random structures
are generated to broadly sample a sub-volume of the total
configuration space, see Figure \ref{figure_landscape}. This
sub-volume is defined by the search parameters. These parameters
include the range of unit cell volumes and shapes, species dependent
minimum distances, structural or molecular units, and symmetry. If
these settings are well chosen, the initial random structures are
‘sensible’ and steer the search to promising regions of the energy
landscape. AIRSS depends on features of the first principles energy
landscape for its effectiveness, in particular its relative
smoothness.\cite{pickard2011ab}

The development of robust first principles codes to calculate the
total energy of extended systems, through periodic boundary
conditions,\cite{kresse1996efficiency,clark2005first,giannozzi2009quantum}
along with databases of accurate
pseudopotentials,\cite{lejaeghere2016reproducibility} has enabled high
throughput computational approaches. One high throughput approach is
to compute properties of structures derived from experimental
databases, such as the
ICSD.\cite{curtarolo2012aflowlib,jain2013commentary} Structure
prediction, and especially AIRSS, also depends on high throughput
computations, with the structures rather generated stochastically.

Density functional theory (DFT) offers a very efficient way to compute
electronic properties from first principles at the quantum mechanical
level,\cite{burke2012perspective} but it remains computationally
expensive in the Kohn-Sham formulation, as single particle
wavefunctions must for optimised for all the electrons in the
system. During the 1980s, as the techniques behind modern DFT codes
were being developed, there was a parallel interest in accelerating
computations using empirical
potentials.\cite{stillinger1985computer,biswas1985interatomic,tersoff1988empirical}
Physically inspired functional forms for the interatomic potentials
were constructed, and the free parameters fit to experimental data, or
small datasets of first principles data.\cite{biswas1985interatomic}
With the advent of high throughput computation, which can rapidly
generate large datasets, these approaches to fit potentials have been
revisited, in the context of machine
learning.\cite{deringer2019machine}

Machine learning has a long history in the materials
sciences.\cite{hkdh1999neural,skinner1995neural} In the 1990s attempts
were made to use neural networks to learn electronic band-structures,
to accelerate Brillouin-Zone integration for Electron Energy Loss
Spectra prediction.\cite{pickard1995ab} Neural networks were also used
to fit complex energy landscapes of isolated
systems,\cite{brown1996combining} density
functionals,\cite{tozer1996exchange} and to predict alloy
properties.\cite{bhadeshia1995impact}

Hampered by a relative lack of data, and the computational costs of
training neural networks, it has taken some time for these approaches
to become ubiquitous. Key to a revitalisation of the application of
machine learning to interatomic potentials has been the work of Behler
and Parrinello,\cite{behler2007generalized} who emphasised the
importance of decomposing the total energy into atomic contributions
for neural network potentials, and Csanyi and
coworkers,\cite{bartok2010gaussian} who introduced the alternative
gaussian approximation potentials. A wide variety of machine learning
potentials are now
available.\cite{bartok2017machine,behler2021four,seko2015first,artrith2016implementation,shapeev2016moment,hajinazar2017stratified,zhang2018deep,benoit2020measuring}
They vary depending on the strategy for assembling the training
data,\cite{skinner1995generating} describing the local
environments,\cite{musil2021physics} and the machine learning model
for regressing the energy landscape.

Structure prediction can be accelerated if the computational cost of
evaluating the energy landscape can be reduced through efficient
approximation.\cite{wu2013adaptive,ouyang2015global,patra2017neural,deringer2018data,tong2018accelerating,kolsbjerg2018neural,thorn2019toward,podryabinkin2019accelerating,hajinazar2021maise}
If that approximation is robust and of sufficiently high quality, for
all, or most, sampled configurations, AIRSS can be attempted. Here the
development of a data derived potential, based on a many-body
environment descriptor and the combination of many small neural
networks, is described. Coupled with an iterative training scheme it
is shown that potentials can be constructed, as needed, for a given
set of search parameters. They are described as ephemeral, as there is
no attempt to build a definitive potential for any given chemical
system, and a new potential can be constructed from scratch at little
cost.

In what follows, the scheme for generating the data derived, ephemeral or
disposable, potential designed for random structure search is
described. It is benchmarked first against a CH$_4$ dataset, then
validated for boron nitride, elemental boron and urea. Finally, in an
true test of the approach, it is used to uncover a complex dense phase
of silane.

\section{A data derivable potential}
\label{addp}

An idea central to the development of potentials is that the total
energy of a collection of $N$ atoms can be decomposed into the individual
contributions of each atom:

\begin{equation}
  E=\sum_i^N E_i.
  \label{totalenergy}
\end{equation}

When combined with the approximation that the energy of each atom,
$E_i$, depends on the environment of that atom within some localised
region, typically a sphere with cutoff radius $r_c$, fast linear scaling
computational schemes are possible.

The energy of each atom, $E_i$, can be further decomposed into terms
that depend on the interactions between increasing numbers of
surrounding atoms:

\begin{equation}
  E_i=E^{(0)}_i+E^{(1)}_i+E^{(2)}_i+E^{(3)}_i+E^{(4)}_i+\cdots.
  \label{atomenergy}
\end{equation}

The zero body term, $E^{(0)}_i$, is typically dropped as it describes
a chemical species independent energy offset, leading to a rigid shift
of the total energy of the system regardless of composition.

The one body term, $E^{(1)}_i$, depends only on the chemical species
of atom $i$. In an elemental system, or one of any fixed composition,
it again leads to an overal rigid shift of the total energy, and can
be ignored. It is vital, however, for the description of compounds
with variable composition.

\subsection{Two body interactions}

The two body term, $E^{(2)}$, is the first that leads to a non-trivial
energy landscape. Physically, it describes the attraction, or
repulsion between pairs of atoms. The earliest potentials applied to
model materials, such as the Lennard-Jones potential, were two body
potentials. The Lennard-Jones potential, with its linear, homogenous,
form compromises between computational efficiency and physical
motivation. This might be contrasted with the inhomogeneous, and
non-linear, Buckingham potential with an exponential term describing
the repulsion between closed electron shells, a $1/r^6$ term
describing attractive dispersion interactions, and a Coulomb term.

Here, we follow the compromise made by
Lennard-Jones, and choose a homogeneous linear potential with the
form:
\begin{equation}
  E^{(2)}_i=\sum_{j\neq i}^N\left(w_1 ^{(2)}f(r_{ij})^{p_1}+w_2 ^{(2)}f(r_{ij})^{p_2}\right),
  \label{twoenergy}
\end{equation}
or
\begin{equation}
  E^{(2)}_i=\sum_{j\neq i}^N\sum_m^2w_m^{(2)} f(r_{ij})^{p_m},
  \label{twoenergy2}
\end{equation}
in the case of two terms (as for the Lennard-Jones potential), and
with a general form:
\begin{equation}
  E^{(2)}_i=\sum_{j\neq i}^N\sum_m^Mw_m ^{(2)}f(r_{ij})^{p_m}.
  \label{twoenergy2}
\end{equation}

The sum is over the $N-1$ other atoms, and over $M$ fixed exponents,
or powers, $p_m$. The linear weights $w_m$ are parameters to be
determined, and the $f(r)$ is a fixed functional form.

For the original Lennard-Jones potential, $f(r)=1/r$, $w_1=1$,
$w_2=-1$, $p_1=12$, and $p_2=6$. Extended Lennard-Jones
potentials\cite{born1940stability} resemble our general form, which
can be written as a scalar product between a weight vector
${\bf w}_{(2)}$, and a vector ${\bf F}^{(2)}_i$, which contains
information about the environment of atom $i$:

\begin{equation}
  E^{(2)}_i=\sum_m^Mw_m ^{(2)}\sum_{j\neq i}^N f(r_{ij})^{p_m}={\bf w}_{(2)}^{\intercal}{\bf F}^{(2)}_i.
  \label{twovector}
\end{equation}

\subsection{Range cutoff}

The Lennard-Jones potential is long ranged, in that there is no
natural cutoff. This range is physically motivated, but it presents
problems for computations of condensed systems. This has long been
recognised, and managed through the imposition of range cutoffs, along
with shifting and adjusting the potential so that it is zero at the
cutoff radius, $r_c$, potentially along with the gradient and higher
derivatives. This is known to have an important impact on the energy
landscape, and indeed the ground state crystal
structures.\cite{partay2017polytypism} Recently, Wang
\emph{et. al.}\cite{wang2020lennard} proposed an alternative to the
Lennard-Jones potential that is appropriately cutoff by construction,
recognising the importance of both computationally efficient and well
defined potentials. Their approach is taken here, and $f(r)$ is
constructed so that it is zero at and beyond $r_c$. There are many
functions which satisfy this condition, but we choose:

\begin{equation}
f(r)=
\begin{dcases}
2(1-r/r_c) & r\le r_c \\
0 & r> r_c.
\end{dcases}
\label{func}
 \end{equation}

\begin{figure}[]
\centering
\includegraphics[width=0.45\textwidth]{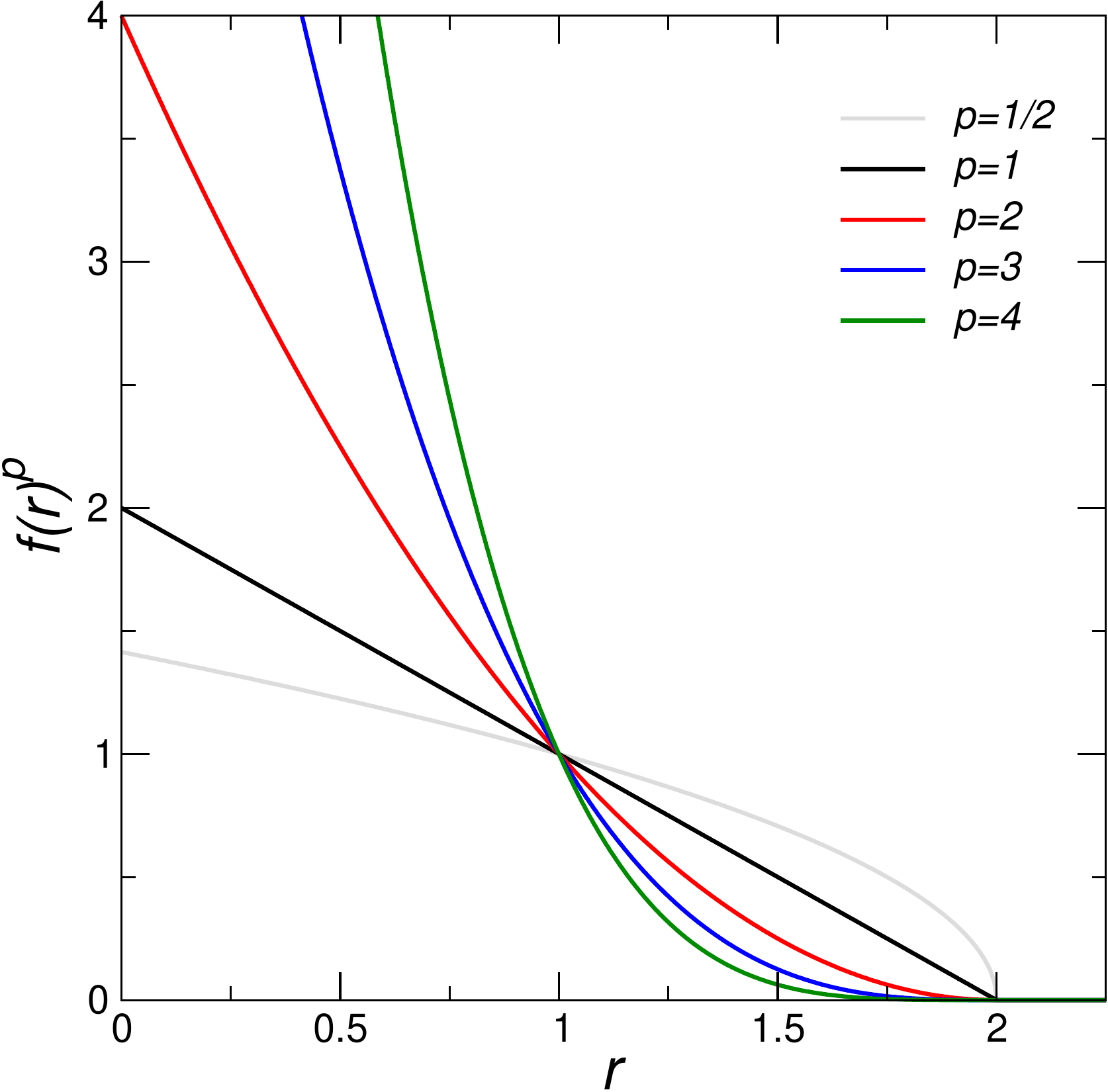}
\caption{The function $f(r)$, defined in Eqn. \ref{func}, raised to a
  range of exponents for the cutoff radius,
  $r_c=2$. }\label{figure_func}
\end{figure}
 
When all the exponents, $p_k$, to which $f(r)$ is raised are two or
greater both the resulting potential, and its gradient, at $r_c$ are
zero, by construction. Higher derivatives can also be forced to be
zero by further increasing the minimum exponent. Exponents that are
less than one (but greater than zero) generate step-like functions,
with steep gradients approaching $r_c$, as shown in
Fig. \ref{figure_func} for $p=1/2$. In what follows all exponents are
chosen to be two or greater.

\subsection{Three body interactions}

Without the careful design of unphysical two body
potentials,\cite{marcotte2013communication} the range of structures
that can be supported in the elements is extremely limited, to those
that are well packed. However, the elements are known to exhibit
extremely rich, and potentially open structures. For example, the
diverse polymorphism in carbon\cite{} and the extremely complex
phosphorous\cite{} and boron\cite{} structures. Contributions are
required to the potential that can distinguish between bond angles in
triplets of atoms. A three body interaction term can achieve this, and
since three distances $r_{ij}$, $r_{ik}$, and $r_{jk}$ uniquely
determine the triangle formed by the three atoms, $i$, $j$, and $k$,
it can be written generally as:

\begin{equation}
  E^{(3)}_i=\sum^N_{j\neq i}\sum^N_{k> j\neq i}V(r_{ij},r_{ik},r_{jk}).
  \label{generalthree}
\end{equation}

The function $V(r_{ij},r_{ik},r_{jk})$ remains to be
parameterised. Consistently with our treatment of the two body
interactions, we write it as a linear, homogeneous, and separable
approximation\cite{biswas1985interatomic}:

\begin{equation}
  E^{(3)}_i=\sum^N_{j\neq i}\sum^N_{k> j \neq i}\sum^M_m\sum^O_ow_{mo}^{(3)}f(r_{ij})^{p_m} f(r_{ik})^{p_m} f(r_{jk})^{q_o}.
  \label{threeatom}
\end{equation}

The individual terms must be invariant to the swapping of the $j$ and $k$
atoms, as is the case in the above by construction. The summation can
be rearranged, as for the two body terms:

\begin{equation}
  E^{(3)}_i=\sum^M_m\sum^O_ow_{mo}^{(3)}\sum^N_{j\neq i}\sum^N_{k> j \neq i} f(r_{ij})^{p_m} f(r_{ik})^{p_m} f(r_{jk})^{q_o},
  \label{threesum}
\end{equation}

and so

\begin{equation}
  E^{(3)}_i=\sum^M_m\sum^O_ow_{mo}^{(3)}F^{(3)}_{i,mo}={\bf w}_{(3)}^{\intercal}{\bf F}^{(3)}_i.
  \label{threevec}
\end{equation}

The three body terms can therefore also be written as a scalar product
between the weight vector ${\bf w}_{(3)}$ and the vector
${\bf F}^{(3)}_i$, which describes the environment around atom $i$,
taking into account three body interactions.

\begin{figure}[]
\centering
\includegraphics[width=0.4\textwidth]{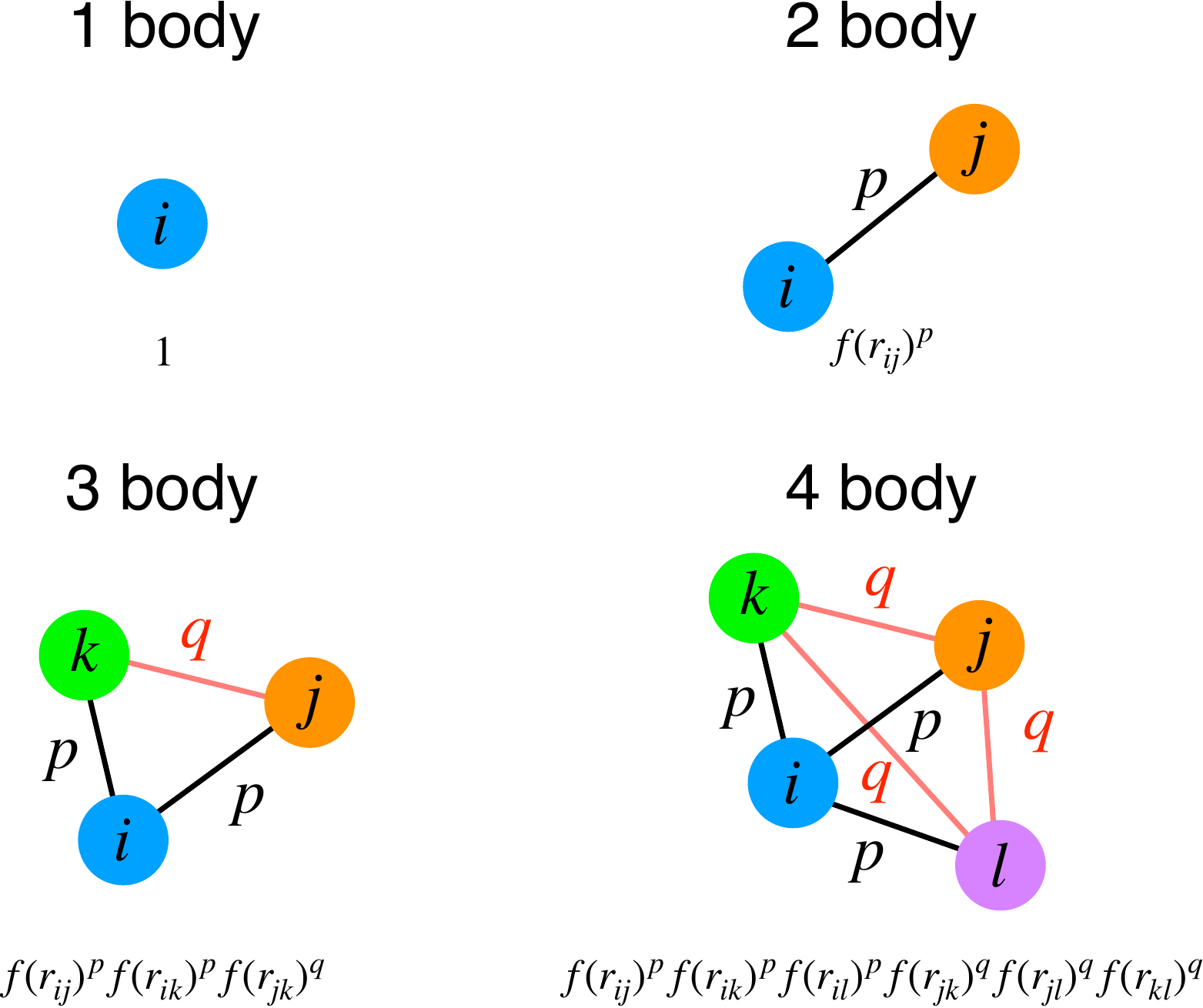}
\caption{Contributions to the environment vectors due to one, two,
  three and four bodies. The exponent $p$ is applied to functions of
the distance from the central atom, $i$, and the exponent $q$ between
the other atoms.}\label{figure_body}
\end{figure}

In principle the construction we have adopted to describe the three
body interactions can be readily extended to four body interactions
(see Fig. \ref{figure_body}) and beyond. However, what follows is
limited to three body potentials througout.

Our construction is related to atomic
body-ordered permutation-invariant polynomials, where our basis is not
complete, but carefully chosen to be computationally efficient and
provide sufficient accuracy.\cite{van2020regularised}

\subsection{Vectorisation  and Multiple species}

For a system containing multiple species the one body contribtion to
the atomic energy, $E_i$, is important.
\begin{equation}
  E^{(1)}_i={\bf w}_{(1)}^{\intercal}{\bf F}^{(1)}_i.
  \label{threevec}
\end{equation}
The one body environment vector, ${\bf F}^{(1)}_i$, has the size of
the total number of species, and assuming full occupancy, one (1) is
added to the n$^{th}$ element if atom $i$ is of species $n$. The two
body environment vector, ${\bf F}^{(2)}_i$, is constructed by
concatenating environment vectors for each of the species pairs. For
example, for two species, A and B:
\begin{equation}
 {\bf F}^{(2)}_i={\bf F}_{{\rm AA},i}^{(2)}\oplus {\bf F}_{{\rm
     AB},i}^{(2)}\oplus {\bf F}_{{\rm BA},i}^{(2)}\oplus {\bf F}_{{\rm
     BB},i}^{(2)}.
 \label{vectorise}
\end{equation}
Note that in the case of full occupancy, and if atom $i$ is of species A
then the second half of the vector will be precisely zero. This leads
to substantial sparsity. The three body environment vector is
similarly constructed from concatenated contributions from triplets of
species, where ${\bf F}_{{\rm ABA},i}^{(3)}$, for example, is
equivalent to ${\bf F}_{{\rm AAB},i}^{(3)}$, and dropped. While it is
not explored further here, this construction is suited to fractional
and mixed occupation.

It is computationally convenient to further concatenate the one, two
and three body environment vectors through the direct sum:
\begin{equation}
 {\bf F}_i={\bf F}_i^{(1)}\oplus {\bf F}_i^{(2)}\oplus {\bf F}_i^{(3)}.
 \label{vectorise}
\end{equation}
This single vector, ${\bf F}_i$, describes the environment of the
atom $i$, considering up to three bodies, and taking atomic species into
account.

\section{Fitting the potential}

Once the environmental (or feature) vectors have been chosen, there
are many possible choices when it comes to the functional form and
fitting procedure. We now describe the scheme selected in this
work. To guide the choices, a number of considerations are made. The
goal is to produce a method that is robust, in that a large fraction
of the structures obtained, on relaxing random sensible structures,
remain sensible and physical. Further, the method should be
computationally rapid. The aim is structure prediction, and the more
time and computational resources spent searching for structures the
better. There should also be a minimum number of parameters, and
reasonable settings that apply to many systems are preferred. The
overall method should demand as little intervention from the user as
feasible.

\subsection{Cost function}
\label{costfn}

The design of the cost function influences the nature of the resulting
fit. While it is common to fit to both the energy landscape itself,
and the forces (and sometimes stresses), which are readily available
within DFT, here we construct a cost function based on total energy
alone:
\begin{equation}
  C=\frac{1}{S}\sum_s\left|\sum^{N_s}_i(E({\bf F}_{s,i})-E_s)\right|^p.
  \label{cost}
\end{equation}
The sum is over the $S$ structures, $s$, in the training data set,
with energies $E_s$ and number of atoms $N_s$. The concatenated
vectors, ${\bf F}_{s,i}$, describing the environment of atom $i$ in
structure $s$ are the input for the function $E({\bf F})$ which
computes the local energy for an atom with environment ${\bf F}$. The
magnitude of the difference between the predicted and target energies
is raised to the power $p$. For $p=2$ the standard least squares cost
function is recovered, whereas for $p=1$, minimising the cost function
reduces the mean absolute error. To deemphasise the impact on the cost
function of a few very poorly predicted local energies (which will
typically be encountered in highly energetic and unphysical structures
far from the low energy structural minima) an intermediate value of
$p=1.25$ is chosen. In principle the individual terms in the cost
function can be weighted.  This is not found to be necessary in the
current scheme.

\subsection{Neural network}
\label{nn}

In Section \ref{addp}, a linear potential was developed from the
environment vectors, ${\bf F}$, and weights ${\bf w}$:
$E_i={\bf w}^{\intercal}{\bf F}_i$. For $p=2$, a closed form for the
weights that minimises the cost function $C$ can be computed. However,
such a potential is limited in the form of the potential energy
surface that can be modelled. Non-linear fits promise to describe more
complex surfaces, but are more challenging to perform. Neural networks
are recognised as a particularly powerful way to carry out general
non-linear fits.\cite{bishop1995neural} They have proven to be
particularly adept for tasks of computational two dimensional image
processing, such as classification. These breakthroughs have been
built on deep (multilayer) neural networks,\cite{bengio2009learning}
with large number of nodes in each layer. The resulting very large
number of weights are optimised through specialist computer codes
running on GPUs.\cite{NEURIPS2019_9015,tensorflow2015-whitepaper} In
this work, in contrast, shallow narrow neural networks are found to be
sufficient, and considerably easier to manage computationally. The
architecture consists of an input layer of the size of the vector
${\bf F}$, a hidden layer with between 5 and 10 nodes, and a single
output node for the predicted atomic energy. The total number of
weights required is modest. Both the inputs and outputs are normalised
on the training data, and a $\tanh$ activation is used between the
input and hidden layer, and a linear activation on output.

\subsection{Levenberg-Marquardt Iteratively Reweighted Least Squares}
\label{lmirls}

Deep neural networks are typically fit (trained) using stochastic
gradient descent,\cite{ruder2016overview} in which gradients are
computed from random subsets (batches) of the training data. Given the
small size of the neural networks employed here, direct minimisation
is more appropriate. General quasi-Newton optimisers empirically did
not perform particularly well for this task, converging slowly to poor
solutions. Given the suitable structure of the cost function, the
powerful Levenberg-Marquardt algorithm can be
used.\cite{levenberg1944method,more1978levenberg} Excellent fits are
reliably obtained in modest numbers of iterations. Although
implemented, geodesic acceleration\cite{transtrum2010nonlinear} was
not observed to significantly improve or speed up the fits in this
case. As originally formulated, the Levenberg-Marquardt algorithm
performs an optimisation of a least squares cost function. For
$p\neq 2$, an approach based on iteratively reweighted least squares
is required.\cite{chartrand2008iteratively} Overfitting is avoided
through early stopping.\cite{prechelt1998early} As the optimisation
progresses the cost of a validation data set, $C_v$, is monitored. If
the validation cost increases for, typically, ten steps the
optimisation is halted and the weights for the minimum $C_v$ are
selected.

\subsection{Non-negative least squares combination}
\label{nnls}

In contrast to linear least square fits, fitting non-linear functions
is a task of non-convex optimisation, leading to a multitude of
potential solutions corresponding to the many local minima of the cost
function depending on the initialisation of the weights. It is
claimed that for neural networks many of these individual solutions
lead to good fits.\cite{dauphin2014identifying} An alternative is to
average a number of fits to produce stabilised ensemble neural
networks.\cite{hansen1990neural,schran2020committee} An attempt was
made to linearly combine multiple fits to minimise the cost function
for the validation data set (to which the neural networks had not been
directly fitted). Extremely low cost functions for both the training
and validation sets can be achieved, given a sufficient number of
individual fits, suggesting that these fits are diverse. However, it
was observed that many of the weights were large and alternating in
sign, and the large costs for the held out testing set implied
overfitting. In any case, such a combination is unphysical. Ideally
one would hope to observe many small positive weights, resulting in an
``adding'' of the individual potentials or fits. To directly enforce
positive weights, non-negative least squares
(NNLS)\cite{chen2010nonnegativity} can be employed. NNLS has the
property of producing sparse solutions, in that the weights are either
positive, or precisely zero. For this application, it is found that
out of, for example, 256 individual neural network fits, around 20 are
selected by the NNLS. The combined NNLS potentials are found to be
considerably more robust than potentials based on single fits. At the
same time they are more computationally efficient than ensemble
averages, automatically discarding any relatively poor individual
fits.

\section{Iterative fitting}
\label{iterative}

\begin{figure}[]
\centering
\includegraphics[width=0.4\textwidth]{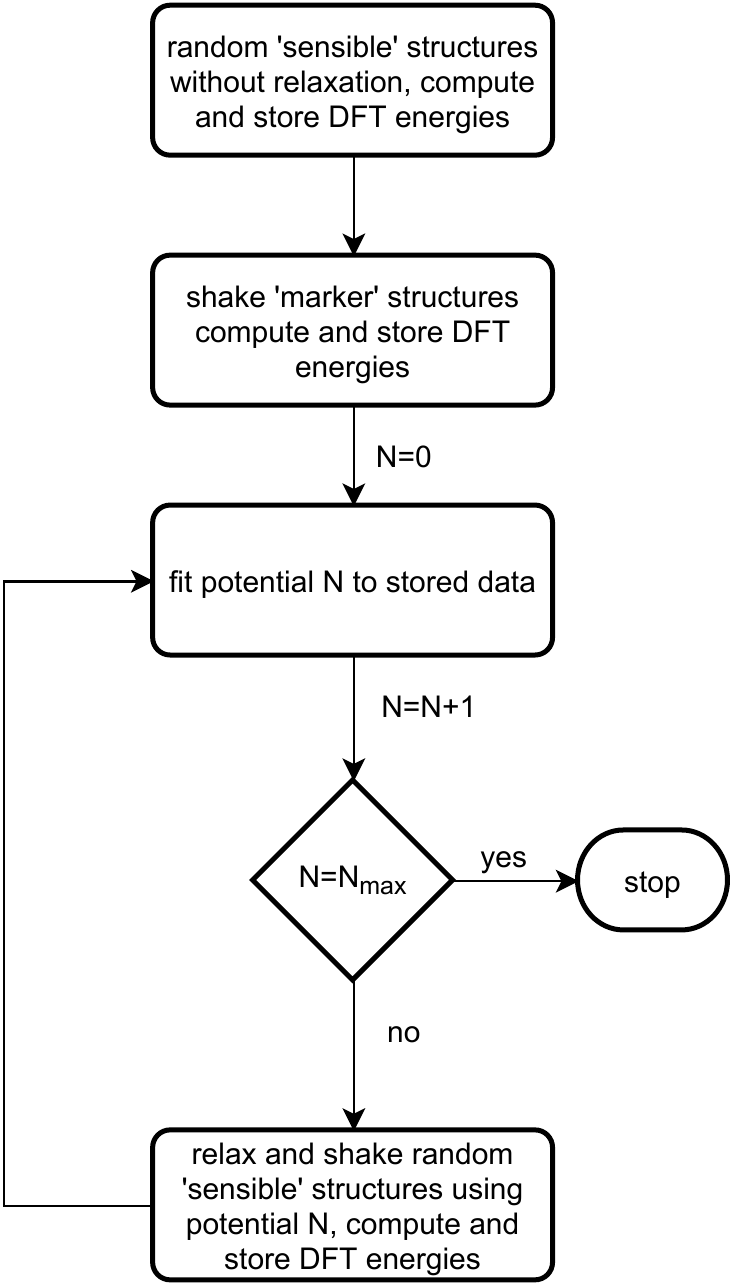}
\caption{A flow diagram outlining the iterative approach to
  fitting. The use of \emph{marker} structures is optional. A typical
  value for N is 5.}\label{figure_flow}
\end{figure}

Closely following the approach developed in
Refs. \onlinecite{deringer2018data} and
\onlinecite{bernstein2019novo}, the fitting is carried out
iteratively, in the manner of the scheme described in
Fig. \ref{figure_flow}. First, random sensible structures are
generated, according to the structure building parameters chosen for
the specific AIRSS search for which the potential will be
used. Without relaxation, the total energies are computed using DFT
and stored along with the structures. These structures will span the
entire region of configuration space accessible consistent with the
biases implied by the AIRSS parameters (for example, unit cell volume
ranges, minimum separations and space groups). Because the structures
are unrelaxed, the typical total energies will be high. These samples
instruct the potential about the high energy regions of the energy
landscape and play an important role in the generation of robust
potentials that are suitable for random search. Without these samples
at high total energy it is likely that the potential will adopt low
and unphysical total energies for these regions of configuration
space. On structural optimisation this can lead to pathological
structures with, for example, extremely close contacts.

The second step is optional, and involves taking so-called ``marker''
structures and applying random small amplitude displacements to their
ionic positions, and lattice vectors. Again, the unrelaxed DFT total
energies are computed and stored. The marker structures are typically
chosen to be known structures in the system of interest. They may be
derived from experiment, or earlier traditional AIRSS searches. Given
that forces and stresses are not present in the cost function, the
role of the shaking of the structures is to provide information about
the gradients of the potential energy landscape. A related approach is
the Taylor expansion method of Ref. \onlinecite{cooper2020efficient}.

At this point, a data set has been generated that is both broadly
representative of the accessible configuration space, and, if marker
structures as selected, of some of the low energy portion of the
energy landscape. The environmental vectors ${\bf F}_{s,i}$ are
computed for all the structures, which are randomly divided into training,
validation and testing subsets in an approximately 80:10:10 ratio. A
potential is then generated using the scheme described in Sections
\ref{nn} to \ref{nnls}.

It is quite likely that the quality of this first fit will not be
particularly good, as monitored through the cost of the held out
testing set, $C_t$. In order to expand the data set, and to ensure the
final potential does not lead to a large number of unphysical low
energy local minima, the following iterative procedure is followed. An
AIRSS calculation is carried out, using the same structure building
parameters and the most recently generated potential, to generate a
number of local minima of the potential energy landscape. These
structures are subjected to a number of random distortions, as for the
marker structures, and the DFT total energies are computed and stored
without relaxation. The combined data set is again randomly split into
training, validation and testing subsets, and a new potential
computed. The next iteration then begins. Either a fixed number of
iterations can be performed, or the procedure halted when the quality
of the fit, as measured by $C_t$, no longer significantly improves.

\section{Implementation}

The implementation consists of a collection of \texttt{OpenMP}
\texttt{Fortran} codes, and \texttt{bash} scripts, assembled into
three separate packages. The \texttt{nn} package is a \texttt{Fortan}
implementation of multilayer neural networks, which is used by the
\texttt{ddp} package to generate the EDDP potentials, and the
\texttt{repose} code which performs variable cell structural
optimisations using a preconditioned\cite{packwood2016universal}
Barzilai-Borwein\cite{barzilai1988two} scheme. The \texttt{ddp}
package consists of several codes. The \texttt{frank} code and
\texttt{franks} script generate the environment vectors for a given
input structure, singly and multiply, respectively. The
\texttt{franks} script exploits the \texttt{parallel}
tool\cite{tange_2020_4045386} to parallelise the environment vector
generation. The \texttt{forge} code performs individual potential
neural network fits, while the \texttt{farm} script manages the high
throughput multiple fits. The \texttt{flock} code combines the
multiple individual fits into a single EDDP using NNLS. The
\texttt{chain} script automates the iterative fitting scheme, and the
\texttt{repose} code is integrated into the GPL2 \texttt{AIRSS}
package.\cite{airsslink} The \texttt{ddp}, \texttt{repose}, and \texttt{nn} packages
are also available under GPL2.\cite{eddplink}

The following examples were computed using a head node with 28 cores
attached to 32 compute nodes, each with 32 cores and accessible by
\texttt{ssh}. Each neural network was trained using 4 \texttt{OpenMP}
cores, permitting 256 fits to be performed in parallel. The CASTEP
plane wave total energy package\cite{clark2005first} is used to
compute the non-spin polarised DFT properties throughout.

\section{Methane molecule}

As a first, and challenging, test we follow
Ref. \onlinecite{pozdnyakov2020incompleteness} and generate a data set
of randomly distorted methane (CH$_4$) molecules. As in
Ref. \onlinecite{pozdnyakov2020incompleteness} the central carbon atom
is fixed, and the four hydrogen atoms are randomly added within a
sphere of radius 3\AA. If any interatomic distance is less than
0.5\AA, the configuration is rejected. The molecule is placed in a
unit cell of side length 10\AA, and the single point total energies
computed using DFT as implemented in the CASTEP
code\cite{clark2005first} with the PBE exchange correlation
functional.\cite{perdew1996generalized} The QC5 on-the-fly
pseudopotentials (\texttt{1|0.9|7|7|9|10(qc=5)} for H, and
\texttt{2|1.4|8|9|10|20:21(qc=5)} for C) are used, with a plane wave
cutoff of 340eV. Generating 10,000 configurations, and dividing them
into training, validation and testing subsets in an approximately
80:10:10 ratio, a three body EDDP is generated five times, with
$r_c=6$, 8 exponents ranging from 2 to 10, and 5 hidden
nodes. Typically, of 256 individual fits, NNLS selects less than
10\%. The best potential of the five resulted in a root mean square
error (RMSE) of 0.13 eV/mol, and the worst 0.18 eV/mol.  Repeating
with 50,000 configurations the best and worst were 0.12 eV/mol and
0.13 eV/mol respectively. The RMSE for 10,000 configurations is
somewhat lower than the best reported in Fig. 4c of
Ref.~\onlinecite{pozdnyakov2020incompleteness}, but the 50,000
configuration result is similar. This suggests that this EDDP, with
its modest number of parameters, performs very well, but the fit does
not improve rapidly with larger datasets. This is an acceptable
compromise for the current application, where low energy candidate
structures will ultimately be relaxed using DFT.

\section{Boron Nitride}

As a first test of the iterative scheme described in Section
\ref{iterative} we explore the construction of a three body EDDP for
boron nitride. Boron nitride adopts a hexagonal layered polymorph as
its most stable form, with the denser tetrahedral cubic polymorph
being metastable. Cubic boron nitride can be synthesised at high
pressures and temperatures.  A hexagonal dense wurztite tetrahedral
structure can also be formed at high pressure.

\subsection{Potential Generation}

\begin{figure}[]
\centering
\includegraphics[width=0.45\textwidth]{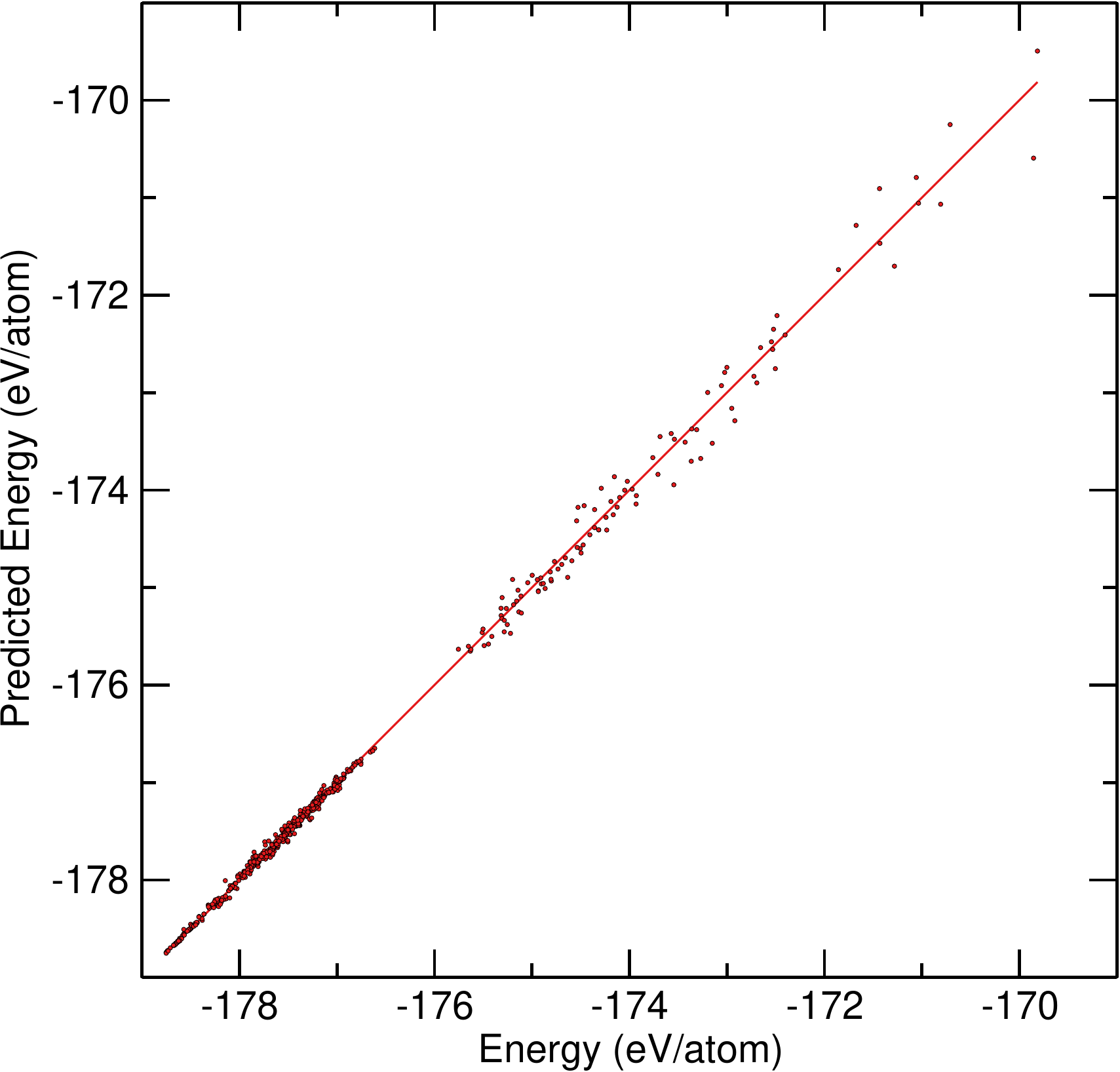}
\caption{The energy per atom predicted by the EDDP plotted against 
  PBE DFT energies for the 650 boron nitride testing configurations. Note that
  despite the relatively large overall RMSE of 86 meV/atom, the error at low
  energies is small,  around 18 meV/atom up to 0.5 eV
  above the ground state, and around 34 meV/atom up to 3 eV.}
\label{figure_energy}
\end{figure}

The EDDP is generated from 4 formula unit (f.u.) boron nitride
structures (8 atoms). The volumes of the unit cells are chosen
randomly and uniformly from 4 to 8 \AA$^3$/atom, no symmetry is
applied, and minimum separations of 1 to 2 \AA~ are randomly
selected. No marker structures are used. 1000 fully random structures
are generated in the first phase, and then 5 cycles of performing
random searching using the current EDDP is performed, generating 100
local minima per cycle. Each of these minima are shaken 10 times, with
an amplitude of 0.02 (AIRSS parameters POSAMP and CELLAMP). The total
energy of each configuration is computed using
CASTEP,\cite{clark2005first} the PBE exchange correlation
functional,\cite{perdew1996generalized} QC5 on-the-fly pseudopotential
(boron definition string \texttt{2|1.4|7|7|9|20:21(qc=5)}, and
nitrogen \texttt{2|1.4|13|15|17|20:21(qc=5)}), with a 440 eV plane
wave cutoff and k-point sampling of 0.05$\times 2\pi$ \AA$^{-1}$. Each
generation of EDDP is constructed using the same parameters. The
cutoff radius, $r_c$, is 3.75\AA, and 4 exponents, ranging from 2 to
10, are used. Non-linear fits (256 in total) are performed with a
neural network with 114 inputs, 5 hidden nodes in a single layer, and a
single output for the predicted atomic energy, and 581 weights in
total. The subsequent NNLS fit to the validation data selects 28
potentials with a non-zero weight. The final EDDP is based on 6495
structures and energies, split into training, validation sets in the
ratio 5196:649:650, and has training, validation and testing RMSE of
42, 55, and 86 meV/atom, respectively. The testing RMSE is
considerably larger then those of the training and validation data
sets. However, as is clear in Figure \ref{figure_energy}, this is the
result of deviations of the predicted energy landscape from the DFT
one only at high energies, and so is benign. The data set contains
structures with energies up to 11.84 eV/atom above the minimum. The
Spearman rank correlation coefficient is above 0.99 for all sets,
suggesting a good ordering of the predicted energies. Including
iteratively building the DFT data set, the EDDP took just 23 minutes
to construct.

\subsection{Structure searches}

Extensive structure searches with the final EDDP and the same
structure generation parameters as used in its construction were
performed for a larger unit cell of 8 f.u. None of the 55,000 fully
relaxed structures contained close contacts. The lowest energy structures
were either layered hexagonal or dense cubic boron nitride, or related
stackings. The energy difference between relaxed hexagonal and cubic
boron nitride is 77.5 meV/atom in PBE DFT, and 79.5 meV/atom using the
EDDP, suggesting that the potential provides an excellent 
ranking at a greatly reduced computational cost. The 55,000 structures
were generated in just 12 minutes using 1024 Intel Xeon Gold 6142
CPU @2.60GHz compute cores. Performing an identical structure search,
using CASTEP for the first principles structural optimisations,
results in 1080 structure over 11.5 hrs. This suggests that searching
using an EDDP is over 250 times faster than DFT for this
application. It should be noted that the EDDP optmisations are
performed to machine precision, while the DFT relaxations are
terminated when the forces and stresses fall below 0.05~eV/\AA~and
0.1~GPa, respectively, which results in far fewer DFT optimisation
steps. The EDDP calculations scale linearly with number of atoms, so
the acceleration for larger systems will grow rapidly. For example, the
computation of the forces and stresses for a 256 atom boron nitride
structure is nearly $10^5$ times faster using the EDDP as compared to
DFT. Should the DFT data have been computed using, for example, a
denser k-point mesh, as would be required for the accurate description
of a metallic system, the acceleration would be larger still.

\subsection{Parameters}

\begin{figure}[]
\centering
\includegraphics[width=0.5\textwidth]{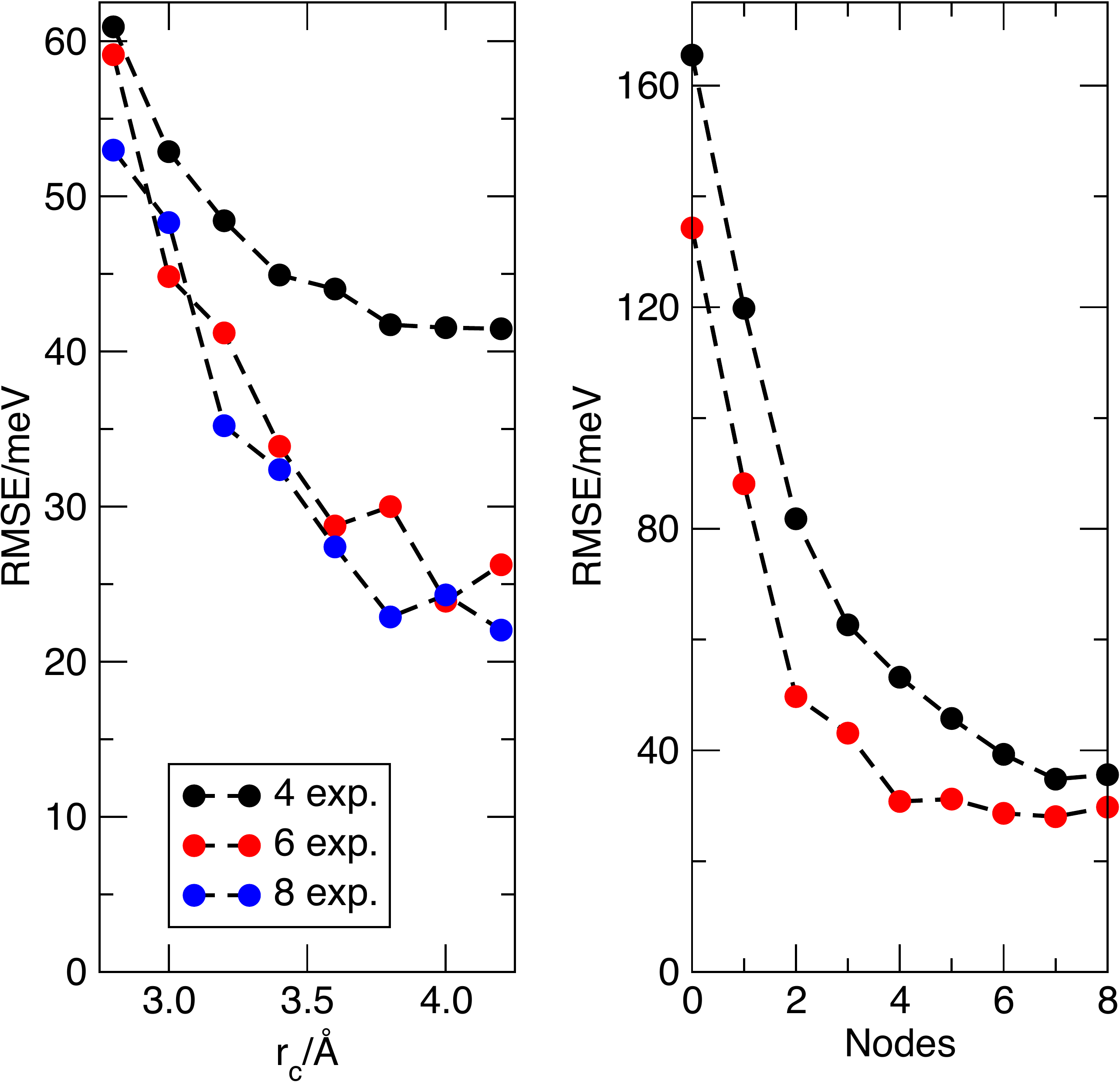}
\caption{The RMSE per atom for EDDPs refit to the interatively
  generated  boron nitride dataset. \emph{Left:} Variation in the fit with the
cutoff radius and number of exponents. \emph{Right:} Variation in the
fit with the number of hidden nodes in the neural networks, and
number of exponents.} \label{figure_parameters}
\end{figure}

The EDDP potential for boron nitride was created without particular
consideration as to the optimal parameters, such as the cutoff radius,
number of exponents, or size of the neural network. The aim is to
perform an accelerated structure search with as little time invested
into potential generation and parameter refinement as
possible. However, it is interesting to investigate how sensitive the
resulting potential might be to the chosen parameters. In Figure
\ref{figure_parameters} the impact of varying the number of exponents,
cutoff radius, and number of hidden nodes, is explored. The previously
iteratively generated data is randomly resplit into training,
validation and testing sets (in the ratio 80:10:10) for each refitting
of the EDDP. It is clear that the 3.75\AA~cutoff radius was a
reasonable choice, but that increasing the number of exponents from 4
to 6 significantly improves the fit. However, increasing further to 8
exponents provides relatively little further improvement, at an
increased computational cost. The fit is also seen to only improve
marginally, if at all, for more than 5 hidden nodes in the neural
networks. Repeating the iterative generation of a three body EDDP with
6 rather that 4 exponents leads to improved training, validation and
testing RMSEs of 26, 38, and 67 meV/atom, respectively. The testing
RMSE is just 20 meV/atom up to 3 eV above the ground state.

\section{Boron}

Elemental boron exhibits extremely complex crystal structures, from
the purely icosahedral $\alpha$-boron, to high pressure
$\gamma$-boron, which consists of icosahedra and dimers which exchange
charge to form an elemental ionic
solid,\cite{oganov2009ionic,zarechnaya2009superhard} and the
exceedingly complex
$\beta$-boron\cite{talley1960new,hoard1970structure}, the structure of
which continues to be studied,\cite{widom2008symmetry} but is thought
to consist of icosahedra and larger defected clusters in a complex
arrangement.\cite{} This structural richness has ensured boron has
played an important role in the development of first principles
crystal structure
prediction.\cite{deringer2019machine,podryabinkin2019accelerating,ahnert2017revealing}
We explore boron as a case study in crystal structure prediction using
EDDPs.

\subsection{Potential Generation}
\label{boronpot}

To reproduce the experience of investigating the boron system without
any prior knowledge, the following procedure is followed. A three body
EDDP is constructed using the iterative scheme detailed above. In the
absence of the knowledge that 12 atom icosahedra are an important
feature of low energy boron structures, the EDDP is generated from
smaller 8 atom unit cells. The volumes of the unit cells are chosen
randomly and uniformly from 3 to 10 \AA$^3$/atom, no symmetry is
applied, and minimum separations of 1 to 3 \AA~ are randomly
selected. In the spirit of a naive search, initially no marker
structures are used. 1000 fully random structures are generated in the
first phase, and then 5 cycles of performing random searching using
the current EDDP is performed, generating 100 local minima per
cycle. Each of these minima are shaken 10 times, with an amplitude of
0.02 (AIRSS parameters POSAMP and CELLAMP). The total energy of each
configuration is computed using CASTEP,\cite{clark2005first} the PBE
exchange correlation functional,\cite{perdew1996generalized} the same
boron QC5 on-the-fly pseudopotential as used for boron nitride, with a
340 eV plane wave cutoff and k-point sampling of 0.05$\times 2\pi$
\AA$^{-1}$. Each generation of EDDP is constructed using the same
parameters. The cutoff radius, $r_c$, is 3.75\AA, and 4 exponents,
ranging from 2 to 10, are used. Non-linear fits (256 in total) are
performed with a neural network with 21 inputs, 5 hidden nodes in a
single layer, and a single output for the predicted atomic energy, and
116 weights in total. The subsequent NNLS fit to the validation data
selects just 15 potentials with a non-zero weight. The final EDDP is
based on 6499 structures and energies, split into training, validation
sets in the ratio 5199:650:650, and has training, validation and
testing RMSE of 52, 52, and 59 meV/atom, respectively. The data set
contains structures with energies up to 11.5 eV/atom above the
minimum. The Spearman rank correlation coefficient is 0.98 for all
sets, suggesting a good ordering of the predicted energies.

\subsection{``Discovery'' of $\alpha$-boron}

As a first test of the EDDP, a random search is performed using the
same structure building parameters as used during the iterative fit,
but with 12 atoms rather than the original 8. Despite the fact that
the training set cannot contain $\alpha$-boron, it is identified as
the most stable structure (once some obviously pathological results,
about 1 in 6000, are removed). How is this possible, given that the
training structures can contain no icosahedra? Examining the most
stable 8 atom structure in the training set (see
Fig. \ref{figure_boron}) it appears that there are hints of
icosahedral fragments in the small cell, which the EDDP is able to
learn, without overfitting, given the relatively inflexible functional
form. It should be noted, unsurprisingly, that this EDDP does not
perfectly reproduce the DFT energy landscape. For instance, it would
be expected to find the $\alpha$-boron structure about 1 in 3000
random samples in a 12 atom unit cell, but using this EDDP it is
reduced to about 1 in 10000 samples. Furthermore, the volume of the
relaxed alpha boron structure differs substantially from the DFT
result, by about 9\%.

\subsection{Structure solution for $\gamma$-boron}

\begin{figure}[]
\centering
\includegraphics[width=0.5\textwidth]{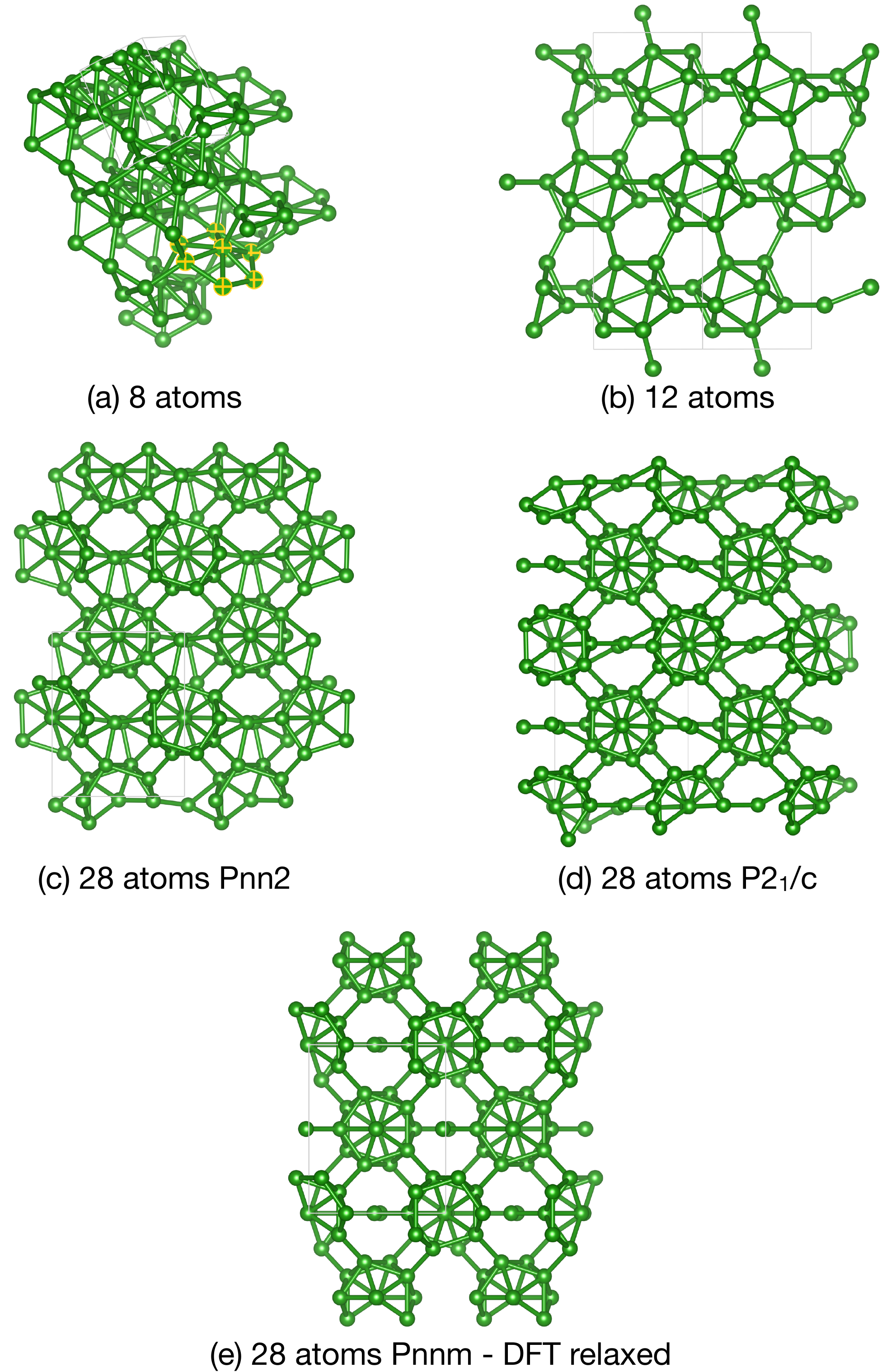}
\caption{Structure (a) is the lowest DFT energy configuration
  contained in the potential training data set for Section
  \ref{boronpot}. A subset of the 8 atoms are highlighted as they
  resemble configuration encountered in icosahedral alpha
  boron. Structure (b) is the result of structure searches using this
  iteratively generated potential, and is that of alpha
  boron. Structure (c) is the result of structure searches in a unit
  cell with shape constrained to that of $\gamma$-boron. Structure (d) is
  the result of structure searches in variable unit cell with no
  imposed symmetry. On relaxation in DFT both the (c) and (d)
  structures become that of $\gamma$-boron, shown in
  (e).} \label{figure_boron}
\end{figure}

A second, more ambitious test, is the solution of the 28 atom gamma
boron structure, from the knowledge of the lattice parameters alone. A
random search, with initial structures with minimum separations of
1.7\AA~and randomly selected space-groups with two to four symmetry
operators, was performed. The fixed unit cell search resulted in about
1 in 3000 obviously pathological structures. The otherwise lowest
energy structures had the Pnn2 space group, a subgroup of Pnnm,
adopted by the $\gamma$-boron structure, see
Fig.\ref{figure_boron}. On inspection the structure appears closely
related to the known $\gamma$-boron structure, and subsequent
structural optimisation of the Pnn2 structure within DFT recovers it
precisely. This result is impressive - it is difficult to conceive
that the 8 atom structures contain obvious hints of the complex
icosahedral/dimer interactions.

\subsection{Free search for $\gamma$-boron}

Next, the challenging task of a symmetry and lattice free search for
the $\gamma$-boron structure is attempted. The EDDP is regenerated
using the $\alpha$-boron structure, which has already been located, as
a marker, which is shaken 500 times. The shake amplitude is increased
to 0.04, the $r_{c}$ to 4.5\AA, the number of exponents to 8 and the
hidden nodes to 10. To increase the chance of encountering
pathological structures during the generation procedure, and to ``dig
deeper'' into the EDDP's energy landscape, on the $N^{\rm th}$ step,
in order to generate a single retained structure, $2^N$ relaxed random
structures are generated, and the lowest energy one selected. Using
this potential, 362,754 structures containing 28 atoms are randomly
generated and relaxed. A dense metastable structure with space group
P2$_1$/c is encountered twice. On inspection the structure appears to
be only a very slight distortion of the $\gamma$-boron structure, and
indeed, on relaxation using CASTEP, it becomes precisely the Pnnm
$\gamma$-boron structure.

\subsection{Structural distortion and potential range}

To test a hypothesis that the observed distortions are due to the
relatively short range of the potentials, a new EDDP is generated,
this time increasing the cutoff to 5.5\AA. The Pnn2 and P2$_1$/c
structures relax directly to the Pnnm structure using this EDDP. There
is clearly a tradeoff between the number of samples that can be
generated, which depends on the computational cost of the potential
used, and the quality of the generated structure. Given that all
important structures in a study will ultimately be relaxed using DFT,
imperfections in computationally cheaper EDDPs can be tolerated in the
pursuit of a more thorough coverage of the energy landscape. However,
care must be taken as a poorly described energy landscape may contain
more local minima, hence be more challenging to search.

Overall, these results for boron suggest that EDDPs are a promising
basis for general random structure prediction tasks.

\section{Urea}

Constructing a fully reactive potential for the entire C-H-N-O
chemical space is expected to present challenges, not least in the
generation and manipulation of suitably large training data sets. In
the spirit if this work, here we generate and apply a three body EDDP
for the specific region of C-H-N-O's configuration space that contains
the urea (CH$_4$N$_2$O) molecule, at around atmospheric and moderate
positive pressures.  Phase transitions in urea (carbamide) under
pressure were first studied by Bridgeman.\cite{} Polymorphism in urea
remains under active investigation, both
experimentally\cite{dziubek2017high} and
computationally.\cite{giberti2015insight,piaggi2018predicting,shang2017crystal}
Here we explore the application of random searching and EDDPs to
identify the low energy polymorphs of urea.

\subsection{Potential Generation}

In the first phase of the iterative construction of the potential for
urea, structures are generated by constructing 10000 randomly shaped
unit cells with volumes from 60 to 80 \AA$^3$/mol, and placing two
urea molecules with random positions and orientations, ensuring that
the molecules are no closer to each other than a randomly selected
distance from 1 to 2 \AA~.  The positions of the atoms in the
molecules are then perturbed by up to 0.3 \AA. The same settings as
for the first potential of boron, Section \ref{boronpot}, are used for
the iterative phases of relaxation and shaking, as well as the final
construction of the potential. The energy of each configuration is
computed using CASTEP,\cite{clark2005first} the QC5 on-the-fly
pseudopotentials (\texttt{2|1.4|13|15|17|20:21(qc=5)} for N, and
\texttt{2|1.5|12|13|15|20:21(qc=5)} for O, and the same definitions
for C and H as for the methane example), and a high plane wave cutoff
of 540 eV. A coarse k-point grid spacing of 0.1$\times 2\pi$
\AA$^{-1}$ was used along with the PBE+TS dispersion corrected
functional.\cite{tkatchenko2009accurate} Of the 256 non-linear neural
network fits, 38 were selected by NNLS. The final potential is based
on 15500 structures and energies, split into training, validation and
testing in the ratio 12400:1550:1550. The training, validation, and
testing RMSE (MAE) is 20.65 (9.99), 27.50 (17.42), and 39.02 (18.76)
meV/atom respectively. The data set contains structures with energies
up to 5.52 eV/atom above the minimum and a Spearman rank correlation
coefficient of 0.999 for all sets, demonstrating an excellent ordering
of the predicted energies.

\subsection{Structure Searches}

\begin{figure}[]
\centering
\includegraphics[width=0.45\textwidth]{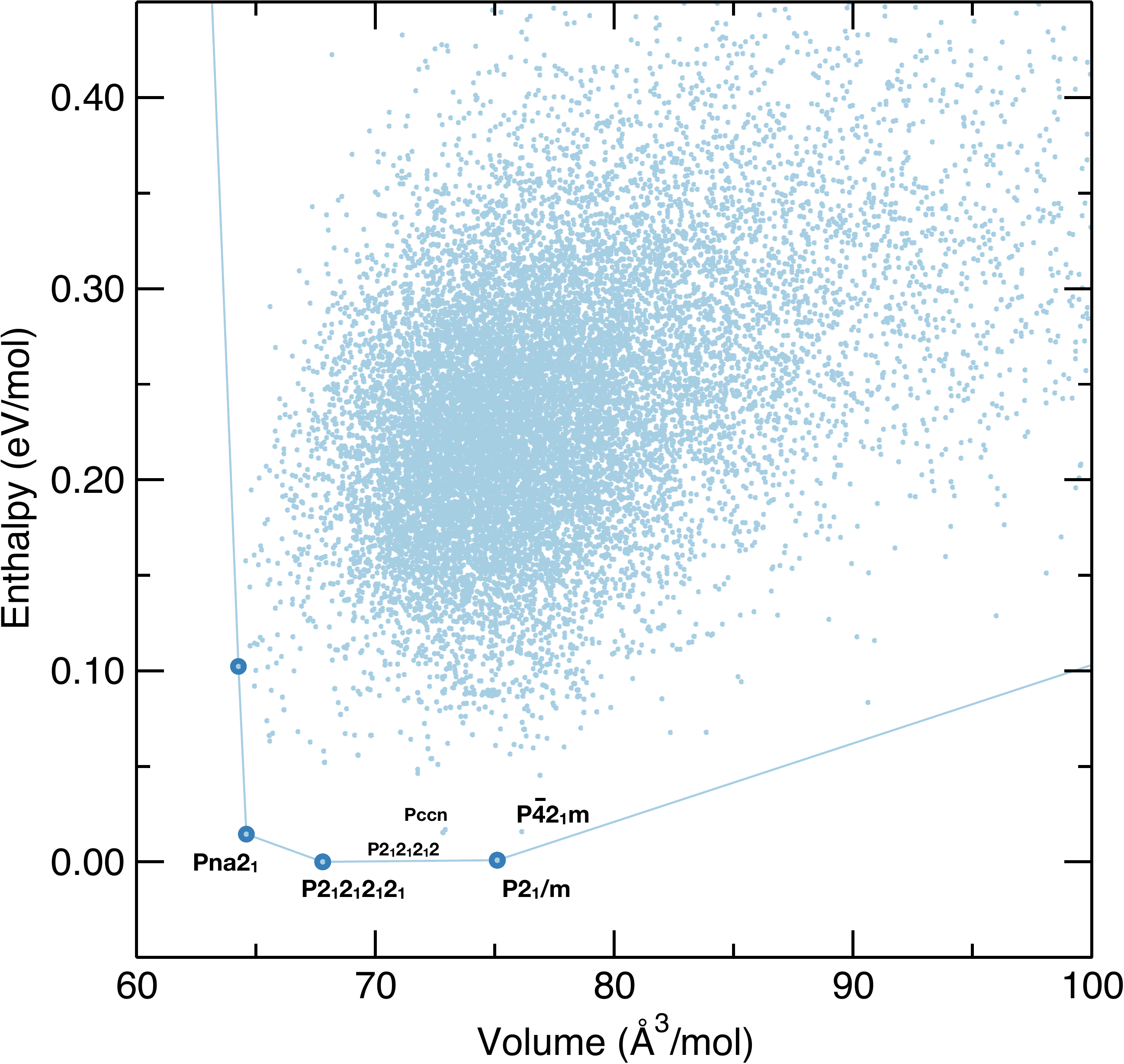}
\caption{Energy versus volume for the 16045 Z=4 urea structures
  relaxed using the EDDP. The fine blue line is the convex hull of the
point, highlighting the structures that might become stable at
positive and negative pressures.} \label{figure_urea}
\end{figure}

Having generated the EDDP for urea using just two molecules per unit cell
(Z=2), it is tested for Z=4. Unit cells with volumes ranging from 60
to 80 \AA$^3$/mol are filled with four molecular units of urea. No
symmetry is used to generate the structures, so in principle
structures with up to Z$'$=4 are accessible. The molecules are placed so
that they do not overlap, with a minimum separation of 2\AA. The
initial structures are relaxed to their nearby local minima 16045
times, generating a diverse set of structures. A scatter plot of the
energy and volume of these structures is shown in Figure
\ref{figure_urea}. 

\begin{figure}[]
\centering
\includegraphics[width=0.4\textwidth]{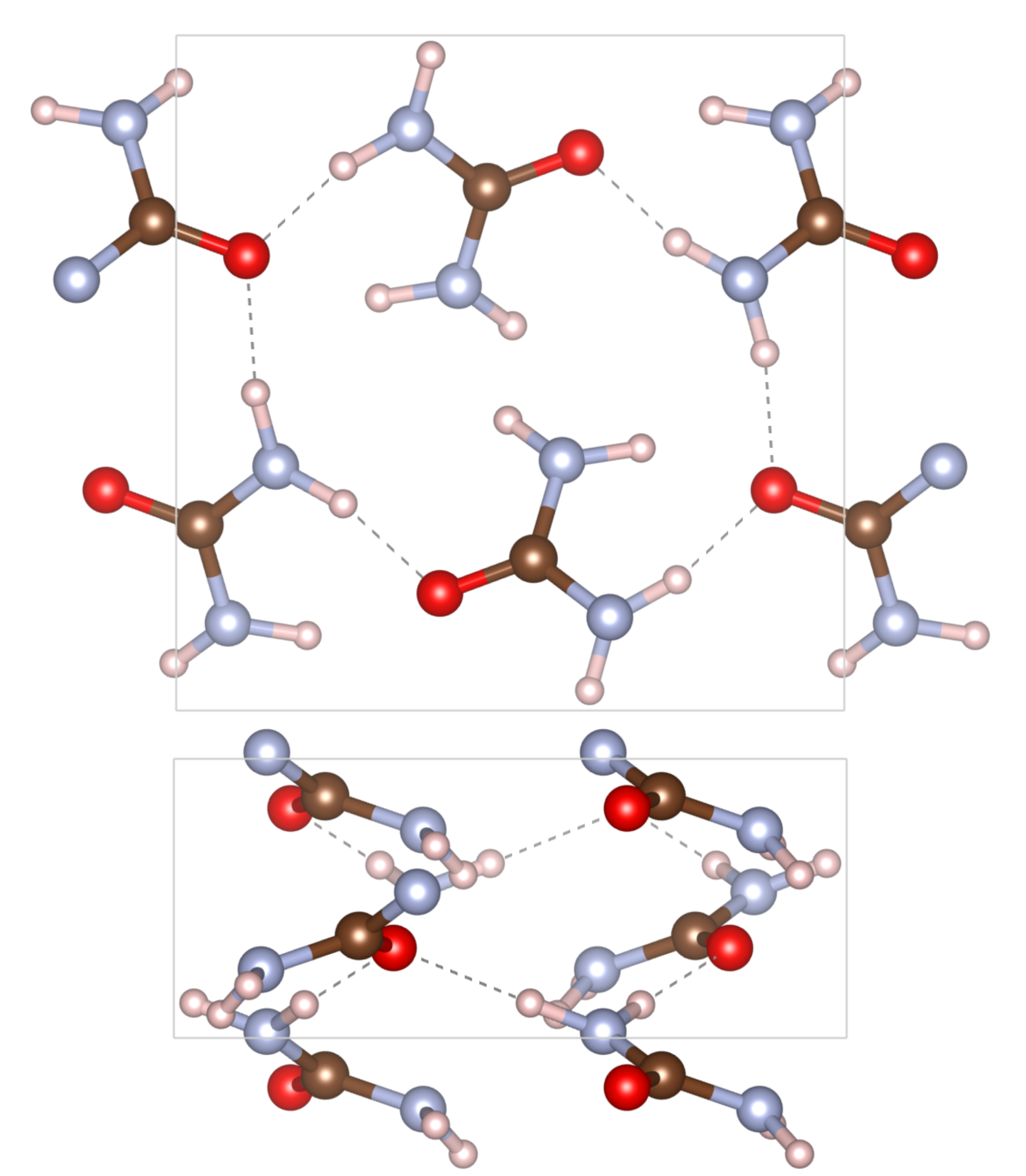}
\caption{The Pna2$_1$ (Z=4) urea structure is energetically
  competitive in all cases, and a candidate high pressure phase of
  urea given its high density.} \label{figure_urea_Pna21}
\end{figure}

The lowest energy structure identified had Z=4 and space group
P2$_1$2$_1$2$_1$. It was located 4 times, and is known as the high
pressure Form III of urea. The ambient pressure P$\bar 4$2$_1$m (Z=2)
form I was located twice, and the high pressure P2$_1$2$_1$2 (Z=2)
form IV was located 18 times. Additional structures with P2$_1$/m
(Z=4), Pna2$_1$ (Z=4) (see Figure \ref{figure_urea_Pna21}), and Pccn
(Z=4) were identified at energies within 40 meV/mol of Form III. To
assess the reliability of the ranking, the structures and energies are
recomputed at both the PBE+TS level (using the same computational
parameters as for the potential generation), and PBE+MBD$^*$ (the
default CASTEP OTFG parameters, a plane wave cutoff of 900 eV and
k-point sampling density of 0.07$\times 2\pi$
\AA$^{-1}$).\cite{ambrosetti2014long} As shown in Table
\ref{table_urea}, in all cases Form III is found to be the lowest
energy structure, with the maximum difference in relative enthalpy
of 40 meV/mol, or 5 meV/atom. It is clear that the EDDP is capable of
resolving differences in energy well below the testing RMSE.

\begin{table}
\begin{tabular}{l|cc|cc|cc}
Space  & \multicolumn{2}{c|}{EDDP} & \multicolumn{2}{c|}{PBE+TS} &
                                                                    \multicolumn{2}{c}{PBE+MBD$^*$} \\
      Group  (Z)             & V/\AA$^3$ & E/meV & V/\AA$^3$& E/meV & V/\AA$^3$ & E/meV\\\hline
  P2$_1$2$_1$2$_1$ (4) & 67.79 & 0 & 68.20 & 0 & 72.15 & 0 \\
  P2$_1$/m (4) & 75.10 & 1 & 74.06 & 24 & 75.31 & 41 \\
  Pna2$_1$ (4) & 64.59 & 14 & 65.37 & 2 & 67.20 & 18 \\
  P2$_1$2$_1$2 (2) & 72.83 & 15 & 70.70 & 17 & 72.87 & 27 \\
  P$\bar 4$2$_1$m (2) & 76.13 & 16 & 71.35 & 13 & 71.86 & 23 \\
  Pccn (4) & 72.93 & 17 & 70.00 & 53 & 70.84 & 55 \\
  
\end{tabular}
\caption{\label{table_urea} Relative energies and volumes (per urea
  molecule) for the low energy structures, evaluated using the EDDP,
  at the PBE+TS level used to construct the potential, and
  PBE+MBD$^*$.}
\end{table}

\section{Application to dense silane}

The earliest published application of first principles random
searching (later referred to as \emph{ab initio} random structure
searching, AIRSS\cite{pickard2011ab}) was to the study of high
pressure polymorphism in silane.\cite{pickard2006high} Feng \emph{et
  al.}\cite{feng2006structures} had proposed silane as a potential
candidate for high temperature conventional superconductivity,using
structures based on chemical intuition and local structural
optimisation using DFT. In Ref. \onlinecite{pickard2006high} random
searches at around 100GPa using two f.u. of SiH$_4$
and just 40 initial configurations uncovered a more stable,
semiconducting, phase of silane with space group {I4$_1$/a}. The
presence of an electronic band-gap postponed any expectation of
superconductivity to higher pressures. Shortly afterwards the I4$_1$/a
structure was encountered
experimentally\cite{eremets2008superconductivity}, and subsequent
theoretical work, exploring larger unit cells of up to 6 f.u.,
identified further candidate structures at both higher and lower
pressures.\cite{martinez2009novel,zhang2015high} Despite refinements
to searching algorithms, and increased computational resources,
structure predictions for binary and ternary compounds are still
typically restricted to relatively small unit cells. Here we revisit
silane, exploiting the computational acceleration afforded by EDDPs to
search in larger unit cells (up to 16 f.u.).

\subsection{Potential Generation}

A three body EDDP was generated using the iterative scheme described
in Section \ref{iterative}. Random unit cells were constructed with
volumes ranging from 5 to 15 \AA/f.u., containing just two f.u. The
minimum separations between the species were randomly chosen to be
between 1 and 2 \AA, and no symmetry was imposed. The total energy of
each configuration is computed using CASTEP,\cite{clark2005first} the
PBE exchange correlation functional,\cite{perdew1996generalized} QC5
on-the-fly pseudopotential (definition strings
\texttt{3|1.8|4|5|5|30:31:32(qc=5)} for Si, and
\texttt{1|0.9|7|7|9|10(qc=5)} for H), with a 340 eV plane wave cutoff
and k-point grid spacing of 0.05$\times 2\pi$ \AA$^{-1}$. The settings
for the iterative scheme, and parameters for the potential, were
identical to those used for boron, with one key difference. The random
searches using each generation of the potential were performed by
minimising the enthalpy at an elevated pressure of 500GPa. This
ensures that the potential will be suitable for high pressure
searches, around this pressure. Non-linear fits (256 in total) are
performed with a neural network with 114 inputs, 5 hidden nodes in a
single layer, and a single output for the predicted energy and 581
weights in total. The subsequent NNLS fit to the validation data set
selects just 23 potentials with a non-zero weight. The final EDDP is
based on 6500 structures and energies, split into training, validation
sets in the ratio 5200:650:650, and has training, validation and
testing RMSE of 9.98, 13.40, and 44.22 meV/atom, respectively. The MAE
error for the testing set is considerably lower, at 10.77 meV/atom,
which is an indication the higher RMSE is the result of a few
structures with significant error. Indeed, the maximum error for the
testing set is 876.02 meV/atom. The data set contains structures with
energies up to 126.4 eV/atom above the minimum. The Spearman rank
correlation coefficient is 0.999 for all sets, suggesting an excellent
ordering of the predicted energies.

\subsection{Structure searches}

Having generated the EDDP suitable for SiH$_4$ at pressures around
500GPa, structures searches may be carried out. As a first test, an
extensive search using the same structure generation parameters as
used for the iterative construction of the EDDP was performed at
500GPa. Any structure that encounters close contacts (by default,
defined at 0.5\AA) during optimisation is rejected. Of the structures
that survive optimisation, the most stable is the C/2c structure
proposed in Ref. \onlinecite{pickard2006high} as the very high
pressure form of SiH$_4$. The 4 f.u. P2$_1$/c structure reported in
Ref. \onlinecite{zhang2015high} is not accessible to a search
restricted to 2 f.u.

The promise of using fast data derived potentials for structure
searching is that much larger systems could be investigated if those
potentials are sufficiently transferable. The challenge of larger
systems is that both each individual structural optimisation is
slower, with each step being more computationally expensive, and the
structural optimisation requiring more of those steps, and that many
more structures must be sampled to ensure the low energy regions of
the energy landscape are adequately explored. Even if the same
structure generation parameters are used for the potential generation
and the search, exploring larger systems is necessarily an
extrapolation. As such, an iteratively generated potential cannot be
expected to result in the precise structures, and energy ordering,
that a DFT search would. However, as we saw in the case of boron
above, the EDDP does appear to offer extrapolation, and
generates appropriate low energy structures. A pragmatic approach is
to simply perform single point energy DFT computations at the end of
each local optimisation using the EDDP. If the EDDP relaxed structures
are reasonably close to what they would be within DFT the ranking
obtained will be reliable, with any poor structures being pushed to
the bottom of the ranking. This is the approach taken here.

We next perform searches at 500GPa with 3 and 4 f.u. of SiH$_4$, again
using the same structure generation parameters, but this time
constructing symmetric initial structures with 2 to 4 symmetry
operations in the primitive cell. The P2$_1$/c structure of
Ref. \onlinecite{zhang2015high} is rapidly recovered, along with the
high pressure C2/c phase of Ref. \onlinecite{pickard2006high}.

\subsection{Identification of complex high pressure phase}

\begin{figure}[]
\centering
\includegraphics[width=0.45\textwidth]{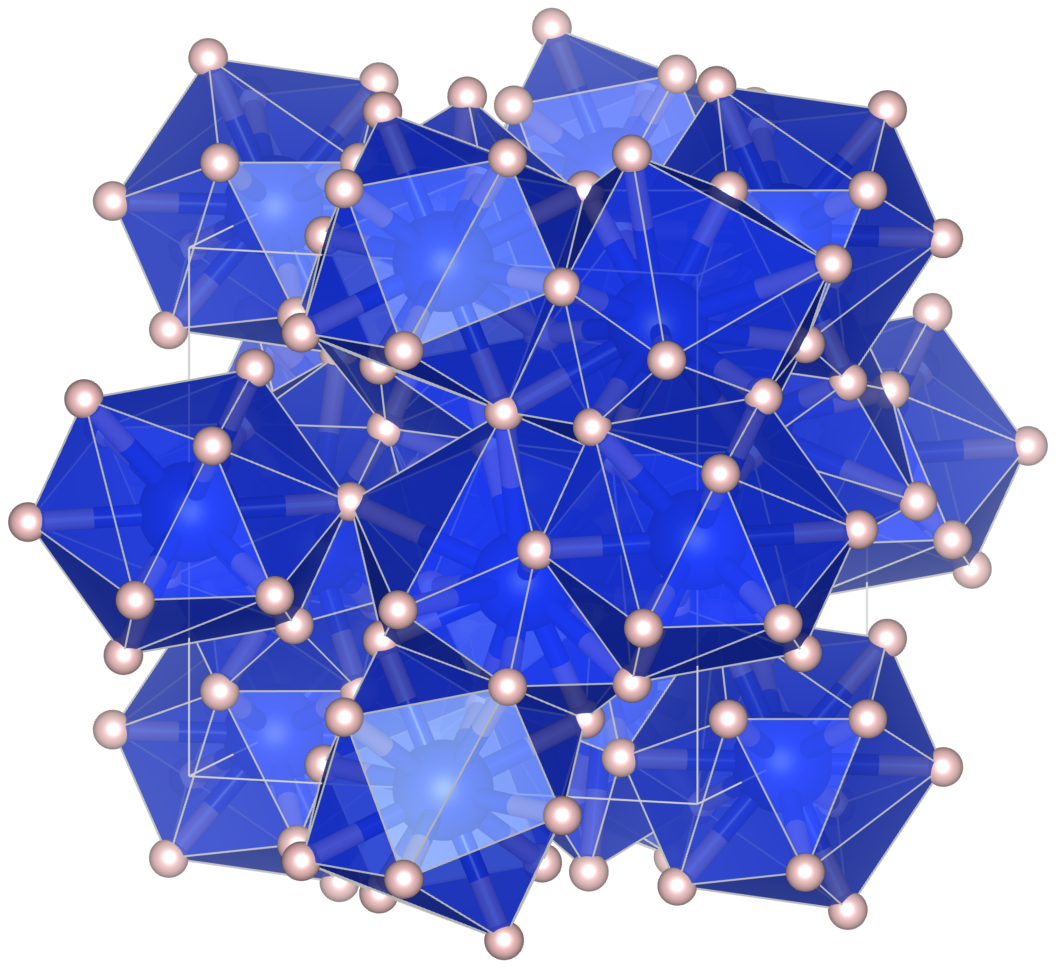}
\caption{A visualisation of the Pa$\bar 3$ structure , created using
  the VESTA package.\cite{momma2011vesta} This complex structure consists of 12
f.u. of SiH$_4$, or 60 atoms, in the primitive unit cell, and does not appear to be
a named structure type.}\label{figure_SiH4_poly}
\end{figure}

Having demonstrated that the potential can recover the theoretically
known high pressure structures of SiH$_4$, its computational
efficiency can be exploited to explore much larger unit cells. A
search at 500GPa is performed with up to 16 f.u. and using between 4 and 12
symmetry operators. A low enthalpy cubic structure with 12 f.u. is
identified, see Figure \ref{figure_SiH4_poly} and Table \ref{table_Pa3}.

\begin{table}
\begin{tabular}{cclllll}
Space group     & {Lattice parameters}            & \multicolumn{4}{c}{Atomic coordinates} \\
                        & {(\AA, $^{\circ}$)}                & \multicolumn{4}{c}{(fractional)}       \\\\\hline\\
Pa$\bar 3$       & $a$=$b$=$c$=4.998                            & Si1 & 0.1168 & 0.1168 & 0.1168 \\
                        & $\alpha$=$\beta$=$\gamma$=90.00 & Si2 & 0.0000 & 0.0000 & 0.5000  \\
                        &                      & H1 & 0.1664 & 0.2236 & 0.3800    \\
                        &                      & H2 & 0.2246 & 0.4858 & 0.3756     
\end{tabular}
\caption{\label{table_Pa3} {Parameters for the Pa$\bar 3$
    structure of SiH$_4$ at 500GPa.}}
\end{table}

\begin{figure}[]
\centering
\includegraphics[width=0.45\textwidth]{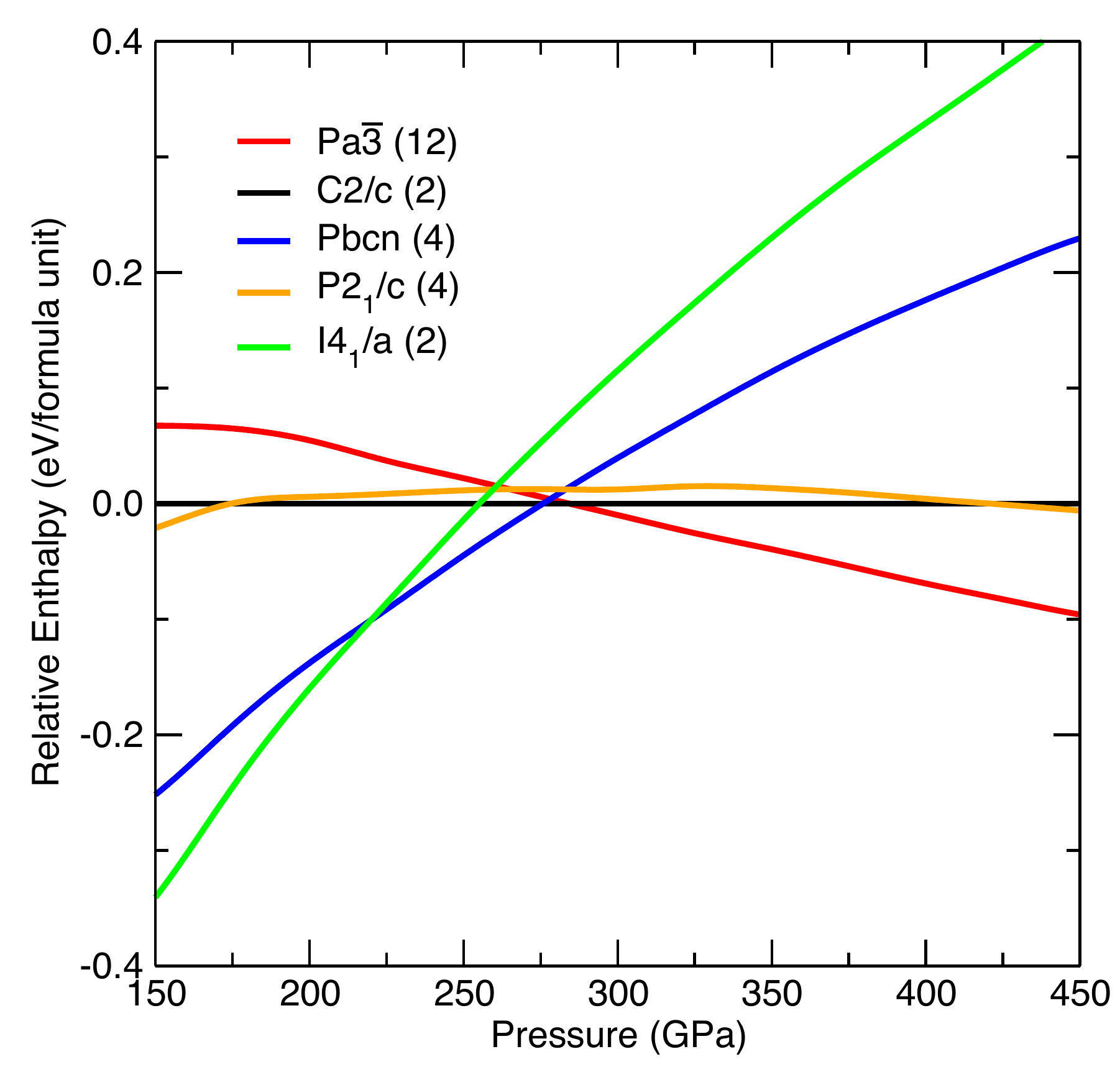}
\caption{Relative PBE DFT enthalpy plotted for a selection of SiH$_4$
  polymorphs. The 60 atom Pa$\bar 3$ structure is increasingly more
  stable than the P2$_1$/c structure above 285GPa, leaving only a
  small window of stability for the C2/c structure from 276 to 285
  GPa.}\label{figure_SiH4_enthalpy}
\end{figure}

\begin{figure}[]
\centering
\includegraphics[width=0.45\textwidth]{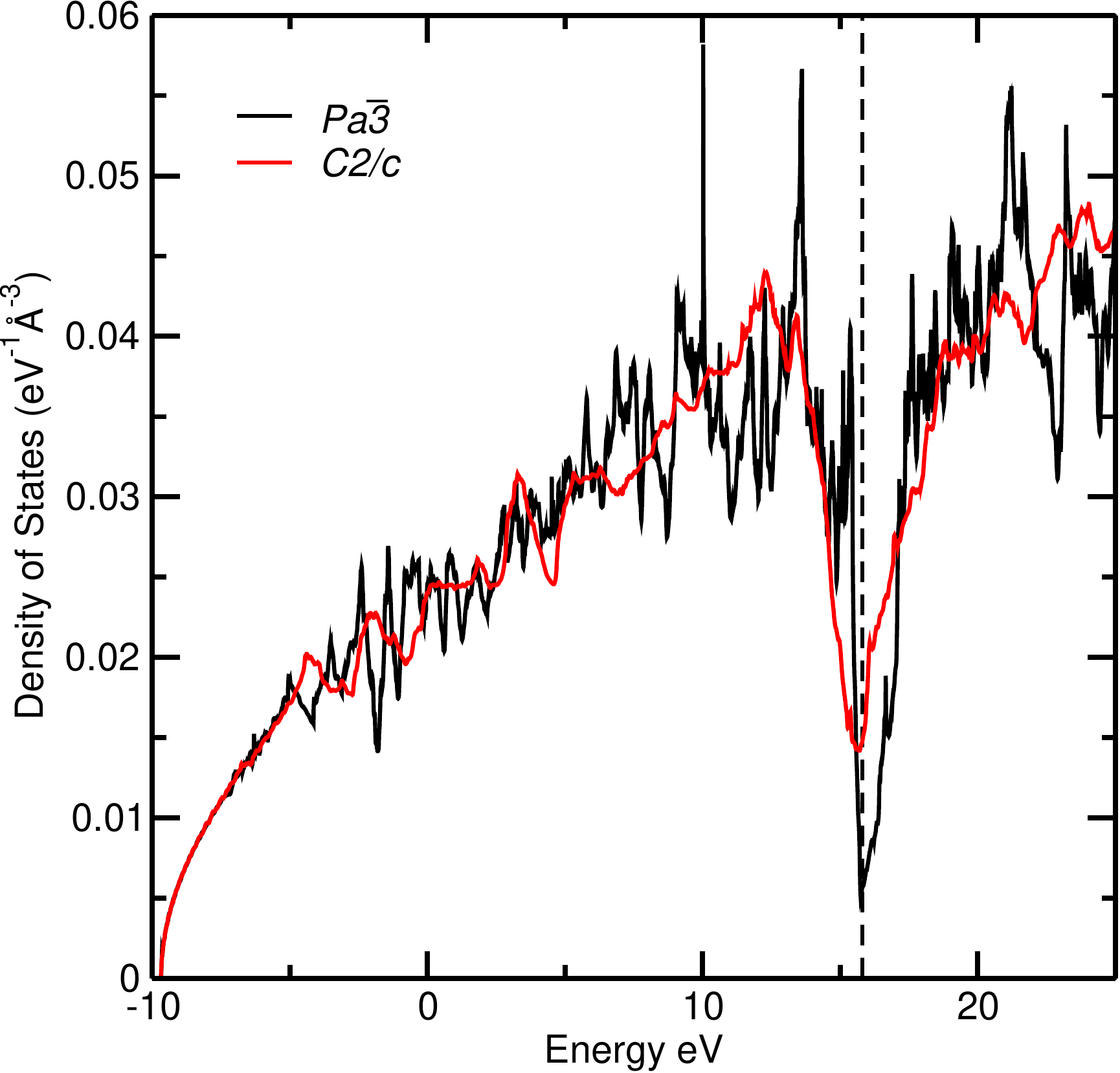}
\caption{The PBE DFT electronic density of states for the Pa$\bar 3$
  and C2/c structures computed at 300GPa. The density of states around
  the Fermi level (vertical dashed line) is considerably lower for the
  Pa$\bar 3$ structure. }\label{figure_sih4_dos}
\end{figure}

This structure adopts the high symmetry Pa$\bar 3$ space group, and is
characterised by two distinct silicon sites, one octahedrally
coordinated by nearest neighbour silicon atoms, and the other
tetrahedrally. To assess its dynamic stability,  a hundred
$3\times3\times3$ supercells of the cubic primitive cell, containing
1620 atoms, were constructed and ``shaken'' with a 0.1 amplitude. On
relaxation with the EDDP all the distorted structure returned to the
60 atom Pa$\bar 3$ space group unit cell. Computing the enthalpy of
this structure, along with those previously reported,
reveals that it has a wide range of stability at the static lattice
level, from 285 GPa upwards using the PBE density functional. Using
the rSCAN\cite{bartok2019regularized} functional it is stable above
305 GPa. It is significantly more dense than the competing phases, and
so its relative stability grows with pressure, see Figure
\ref{figure_SiH4_enthalpy}. The enthalpy curves were computed using
CASTEP, a more accurate potential for hydrogen
(\texttt{1|0.6|13|15|17|10(qc=8)}) and an increased plane wave cutoff
of 700 eV.
The electronic density of states (eDOS) for the Pa$\bar 3$ and C2/c
structures are reported in Figure \ref{figure_sih4_dos}. They were
computed\cite{morris2014optados} with the same settings as for the
enthalpy curves, but with a finer k-point grid spacing of
0.01$\times 2\pi$ \AA$^{-1}$. The eDOS at the Fermi level for the
Pa$\bar 3$ structure is considerably lower than for the C2/c structure
at 300 GPa, which can be attributed to its greater
stability. Furthermore, without performing extremely costly density
functional perturbation theory computations of T$_{\rm c}$ it is
expected that this reduced eDOS would lower the prospects for high
temperature superconductivity in silane at these pressures. Given that
silane has been extensively studied theoretically, the emergence of
such an important, and large unit cell, structure should inform our
confidence in the status of our knowledge of the dense hydrides. It is
very likely that more extensive searches for the dense binary
hydrides, in large unit cells, will reveal a significant revision of
our knowledge of these candidate high temperature
superconductors.\cite{pickard2020superconducting}

\section{Discussion}

First-principles methods owe their flexibility and applicability to
databases of high-quality pseudopotentials, which allow arbitrary
chemical systems to be explored. The CASTEP code\cite{clark2005first} is unique in its on-the-fly
pseudopotential methodology, where the pseudopotential is generated as
needed, and consistently with the density functional chosen. This has
opened the door to structure predictions at extreme densities, with
small core potentials being generated as needed, and independently of
the provided databases.

Here, the same flexibility is introduced to data derived potentials,
which are generated specifically for the structure building
parameters, and pressures, that will be used for each search. These
potentials are ephemeral, in the sense that the next search performed
will likely require a new, bespoke, potential. The ease and robustness
of the scheme described makes this possible.

Random structure search is a challenging application of data derived
potentials. It is very difficult to construct potentials that are
stable across the entire space of possible inputs, or
configurations. The initial random structures are extremely diverse,
exploring many different regions of configuration space. Constructing
the EDDPs from these diverse structures, generated from a given set of
structure building parameters, is essential to ensure robustness.

For any finite training dataset, some failures are to be expected in
an extended sampling of configuration space. A typical pathological
behaviour is the encounter of very close contacts during structural
optimisation or evolution. This could cause severe problems in a
lengthy molecular dynamics simulations. However, during a random
structure search such configurations may simply be rejected. A very
similar situation is encountered in first-principles structure
searches – for heavier elements, overlapping pseudopotentials cores
can lead to problems in the calculation of the electronic structure,
and common practice is to reject those configurations.

The pioneering work of Behler, and Csanyi, who introduced neural
network, and Gaussian process based atomic potentials respectively,
which can be fit to extensive databases of first-principles data, has
led to an explosion of alternative schemes based on their key
insights. It is worth reflecting on the justification of introducing
yet another. In some sense, it is inevitable – there are countless valid
approaches to the fitting of high dimensional functions, and while any
scheme will share commonalities with the others in use, the details
may differ, depending on the intended application. While the
electronic structure community has coalesced around a few, very
complex, computer codes, the relative simplicity of data derived
potentials is likely to favour persistent diversity. In this case a
scheme has been designed for random structure search.

The functional form for the EDDP has its origin in an earlier attempt to
develop a few-parameter model 3-body potential that could describe the
rich structure of the elements, going beyond simple close
packing. Starting with the Lennard-Jones potential, this original model
potential was written as follows:

\begin{equation}
  E_i=\sum_{i\neq
    j}\left(\frac{A}{r_{ij}^{12}}-\frac{B}{r_{ij}^{6}}\right)+\sum_{j\neq
    i}\sum_{k>j\neq i}\frac{C}{r_{ij}^n r_{ik}^n r_{jk}^m}.
  \label{model}
\end{equation}

By manually adjusting the parameters, $A$, $B$, $C$, $n$, and $m$, and
performing random searches for each choice, it was found to be
possible to navigate the space of possible elemental structures, from
close packed, to the diamond lattice, and even the icosahedral
$\alpha$-boron structure. Exploring the properties of the simplified
potential described in Eqn. \ref{model} would be a fruitful topic of
further investigation.

\section{Conclusion}

Fitting of potentials to data generated across the whole accessible
energy landscape ensures that the benign properties of the
first-principles energy landscape are retained, and random search can
be successfully performed. The computational simplicity of the form of
the potential ensures that these searches are much accelerated
compared to a purely first-principles approach. Close attention has
been paid to develop a bespoke scheme that complements the
computational workflow of structure search.

It has been shown that the EDDP potentials can be fit to
first-principles data derived from much smaller unit cells than are
typically chosen for training. These potentials can be used to
discover novel structural features in much larger unit cells. For
example, a potential trained using unit cells containing just eight
boron atoms was used to generate approximations to the 12-atom
icosahedral alpha-boron structure, and the 28 atom gamma-boron. This
extrapolation to larger unit cell sizes is essential if these
potentials are to be successfully used to accelerate structure
prediction.

EDDPs have been used to revisit the high-pressure phase diagram of silane,
uncovering a large (60-atom) unit cell structure that is considerably
more stable at high pressures than those currently known. This
structure had been overlooked, despite extensive investigation using
both random search and evolutionary approaches. This is strong
evidence that EDDPs are a powerful tool for the thorough exploration
of structure space. At the same time, it suggests that many of the
systems that have been explored using first-principles structure
prediction should be revisited.

\begin{acknowledgments}
  CJP is supported by the EPSRC through grants EP/P022596/1, and
  EP/S021981/1, and thanks Gabor Csanyi, Chuck Witt and Lewis Conway
  for discussions, and further thanks Gabor Csanyi for his careful
  reading of the manuscript. 
\end{acknowledgments}


%

\end{document}